\documentclass[twocolumn,twocolappendix]{aastex63}

\usepackage{chngcntr}
\usepackage{enumitem}
\usepackage{amsmath}
\usepackage{txfonts}
\usepackage{xcolor}
\usepackage{graphicx}
\usepackage{gensymb}
\usepackage{booktabs}
\usepackage{bm}
\usepackage{lipsum}
\usepackage{fnpct}

\AtBeginDocument{\mathcode`v=\varv}

\providecommand{\sorthelp}[1]{} 

\newcommand{\HI}{\ifmmode \mathrm{\ion{H}{1}} \else \ion{H}{1} \fi}
\newcommand{\CII}{\ifmmode \mathrm{\ion{C}{2}} \else \ion{C}{2} \fi}
\newcommand{\nh}{\ifmmode N_{{\mathrm{H}} \, \mathrm{I}} \else $N_{{\mathrm{H}} \, \mathrm{I}}$\fi} 
\newcommand{\ROHSA}{{\tt ROHSA}}

\newcommand{\norm}[1]{\left\lVert#1\right\rVert}

\newcommand\ab{\bm{a}}
\newcommand\mb{\bm{m}}
\newcommand\rb{\bm{r}}

\newcommand\mub{\bm{\mu}}
\newcommand\sigmab{\bm{\sigma}}
\newcommand\thetab{\bm{\theta}}

\newcommand{\Trb}{T_{\mathrm b}}
\newcommand{\Trc}{T_{\mathrm c}}
\newcommand{\Trn}{T_{\mathrm n}}
\newcommand{\Trs}{T_{\mathrm s}}
\newcommand{\Tsys}{T_{\mathrm sys}}

\newcommand{\DFG}{DF} 
\newcommand{\UMG}{UM} 

\newcommand{\Planck}{\textit{Planck}}

\def\GHz{\ifmmode $\,GHz$\else \,GHz\fi}
\def\MJysr{\ifmmode \,$MJy\,sr\mo$\else \,MJy\,sr\mo\fi}
\def\microns{\ifmmode \,\mu$m$\else \,$\mu$m\fi}

\def\kms{\ifmmode $\,km\,s$^{-1}\else \,km\,s$^{-1}$\fi}

\defcitealias{wolfire_neutral_2003}{W03}
\defcitealias{wolfire_neutral_1995}{W95a}
\defcitealias{wolfire_multiphase_1995}{W95b}
\defcitealias{marchal_2019}{M19}
\defcitealias{leike_2020}{L20}
\defcitealias{blagrave_dhigls:_2017}{B17}

\received{May 8, 2022}
\revised{August 4, 2022}
\accepted{August 19, 2022}
\published{-- --, 2022}
\submitjournal{ApJ}

\shorttitle{Thermal condensation of diffuse \HI\ in the NCPL}
\shortauthors{Taank, Marchal, Martin, and Vujeva}

\graphicspath{{./}{figures/}}

\begin{document}

\title{Mapping the thermal condensation of diffuse \HI in the North Celestial Pole Loop}

\correspondingauthor{Mukesh Taank}
\email{mtaank@uoguelph.ca}

\author[0000-0001-8461-5552]{Mukesh Taank}
\affiliation{Department of Mathematics and Statistics, University of Guelph, 50 Stone Road E., Guelph, ON N1G 2W1, Canada}
\affiliation{Canadian Institute for Theoretical Astrophysics, University of Toronto, 60 St. George Street, Toronto, ON M5S 3H8, Canada}

\author[0000-0002-5501-232X]{Antoine Marchal}
\affiliation{Canadian Institute for Theoretical Astrophysics, University of Toronto, 60 St. George Street, Toronto, ON M5S 3H8, Canada}

\author[0000-0002-5236-3896]{Peter G. Martin}
\affiliation{Canadian Institute for Theoretical Astrophysics, University of Toronto, 60 St. George Street, Toronto, ON M5S 3H8, Canada}

\author[0000-0001-7697-8361]{Luka Vujeva}
\affiliation{Canadian Institute for Theoretical Astrophysics, University of Toronto, 60 St. George Street, Toronto, ON M5S 3H8, Canada}

\begin{abstract}
    The North Celestial Pole Loop (NCPL) provides a unique laboratory for studying the early stages of star formation, in particular the condensation of the neutral interstellar medium (ISM). Understanding the physical properties that control the evolution of its contents is key to uncovering the origin of the NCPL.
    %
   %
    %
    Archival data from the NCPL region of the GHIGLS 21\,cm line survey (9\farcm4) are used to map its multiphase content with {\tt ROHSA}, a Gaussian decomposition tool that includes spatial regularization. Column density and mass fraction maps of each phase were extracted along with their uncertainties.
    Archival data from the DHIGLS 21\,cm (1') survey are used to further probe the multiphase content of the NCPL. %
    We have identified four spatially (and dynamically) coherent components in the NCPL, one of which is a remarkably well-defined arch moving at about 14\,\kms\ away from us that could be a relic of the 
    large scale organized dynamical process at the origin of the phase transition. The cold and lukewarm phases together dominate the mass content of the neutral gas along the loop. Using absorption measurements, we find that the cold phase exhibits slightly supersonic turbulence.
\end{abstract}


\section{Introduction}
\label{sec:intro}

The possible coexistence of two thermal phases in pressure equilibrium in the neutral atomic interstellar medium was described by \citet{field_thermal_1965} and \citet{field_cosmic-ray_1969}, but being an equilibrium theory the amount of gas in each phase is indeterminate. 
It is expected that the condensation process from the warm neutral medium (WNM) requires perturbation by a dynamical event (e.g., colliding flows, expanding shells, turbulence) to trigger thermal instability, resulting in cooler and denser gas, the cold neutral medium (CNM). Indeed that is borne out by simulations \citep{hennebelle_dynamical_1999,seifried_2011,saury_2014,bellomi_2020}, but the fractional amount of CNM arising is dependent on the details \citep{marchal_2021a,marchal_2021b}.
This condensation can be an evolutionary precursor to molecular gas \citep{rybarczyk_2022} and star formation and so is of great interest and important in the interstellar medium (ISM).

The North Celestial Pole Loop 
\citep[NCPL, ][]{heiles_1984,heiles_1989,meyer91}
provides a favorable environment in which to study the phase transition and its outcome empirically: finding gas and dust in a prominent feature so far from the Galactic mid-plane argues for a large scale organized dynamical process \citep{mandm22}, and being at intermediate latitude the line of sight is not heavily contaminated by foreground and background emission.

Early observational evidence for multiphase gas in the NCPL was reported by \citet{heiles_1989} in his
study of the Zeeman effect for the 21\,cm line.
Although a one-Gaussian simultaneous fit to the Stokes parameter $I$ and $V$ spectra produced a fairly good solution, he found that a two-Gaussian model was needed to reproduce broad tails seen in the spectra. One component is narrow like CNM, while the other is broader, reminiscent of the WNM. 
However, he also found that the broad component was typically narrower than expected for warm gas at 8000\,K and concluded that the ``WNM is not \textit{all} at the same temperature" and that this might be the signature of gas in the thermally unstable regime.
Focusing on Spider region at the ``top" of the loop (in Galactic coordinates, see Fig.~\ref{fig:T_b_full_cube_map}), \citet[][hereafter B17]{blagrave_dhigls:_2017} found evidence that the intricate structures seen in \HI\ (and in thermal dust emission) arise from the CNM phase. 

In this paper, we use high-resolution \HI\ data and the advanced multiphase analysis tool {\tt ROHSA} \cite[][hereafter M19]{marchal_2019} to quantify the multiphase properties of \HI\ gas in the NCPL, thus probing the impact of the inferred dynamical process on the structure of the diffuse gas and providing additional clues about the dynamics. 
The long-standing question of the origin of the loop itself is taken up separately \citep{mandm22}. Together, these works should motivate and constrain future simulations.

The paper is organized as follows. 
In Section~\ref{sec:data} we present the \HI\ data used in this work.  
Section~\ref{sec:methods} describes the Gaussian decomposition performed to model the \HI\ spectra
and Section~\ref{sec:id-phases} provides a quantification of the thermal phases identified in the NCPL. 
Further evidence for cold gas from absorption lines is presented in Section~\ref{sec:absorption1}.
Finally, a discussion and summary are provided in Sections~\ref{sec:discussion} and \ref{sec:summary}, respectively.

\section{Data} \label{sec:data}

We used the GHIGLS\footnote{GBT \HI Intermediate Galactic Latitude Survey: \url{https://www.cita.utoronto.ca/GHIGLS/}} spectra of the \HI\ 21-cm line brightness temperature, $T_{\rm b}$, from the Green Bank
Telescope (GBT) \citep{martin_ghigls:_2015}, to probe the structure of neutral gas in the NCPL.
Key characteristics of the relevant data cubes are as follows \citep{boothroyd2011}:
spatial resolution and pixel size 9\farcm4 and 3\farcm5, respectively; velocity resolution and channel spacing  $1.0\,$\kms\ and $0.8\,$\kms, respectively; velocities with respect to the Local Standard of Rest (LSR) in a range at least $-310 < v < +210\,$\kms. 

\begin{figure}[!t]
    \centering
    \includegraphics[width=\linewidth]{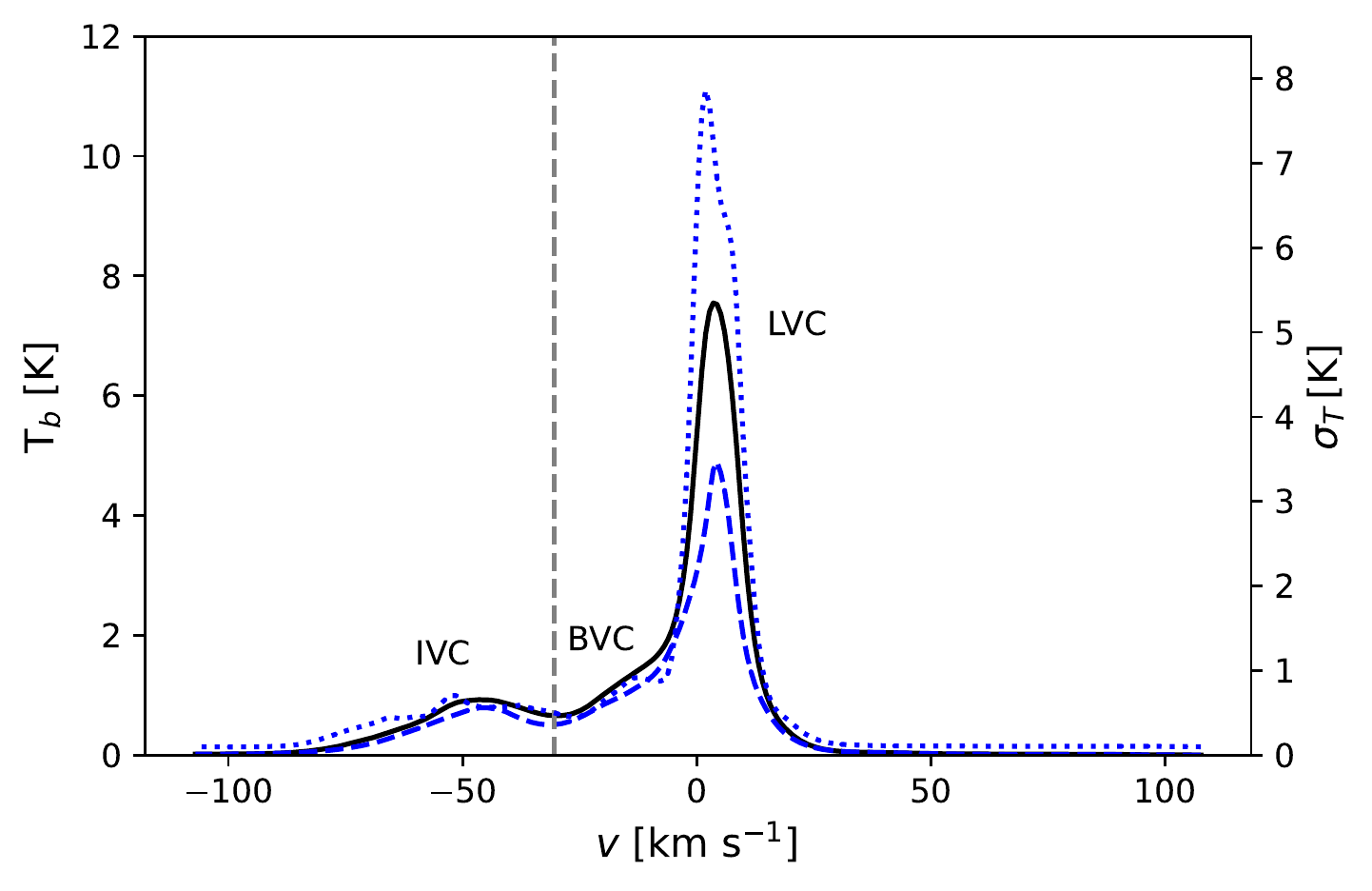}
    \caption{Mean \HI spectrum of the region within the GHIGLS NCPL mosaic analysed, the white box in Figure~\ref{fig:T_b_full_cube_map}, showing relative positions of the IVC, BVC, and LVC. The black and dashed blue lines are the mean and median spectra, respectively (left y-axis). The dotted blue line is the standard deviation spectrum (right y-axis). Our analysis focuses on velocity channels to the right of the dashed grey line.
    }
    \label{fig:avgspec}
\end{figure}

\begin{figure*}[!t]
    \centering
    \includegraphics[width=0.85\linewidth]{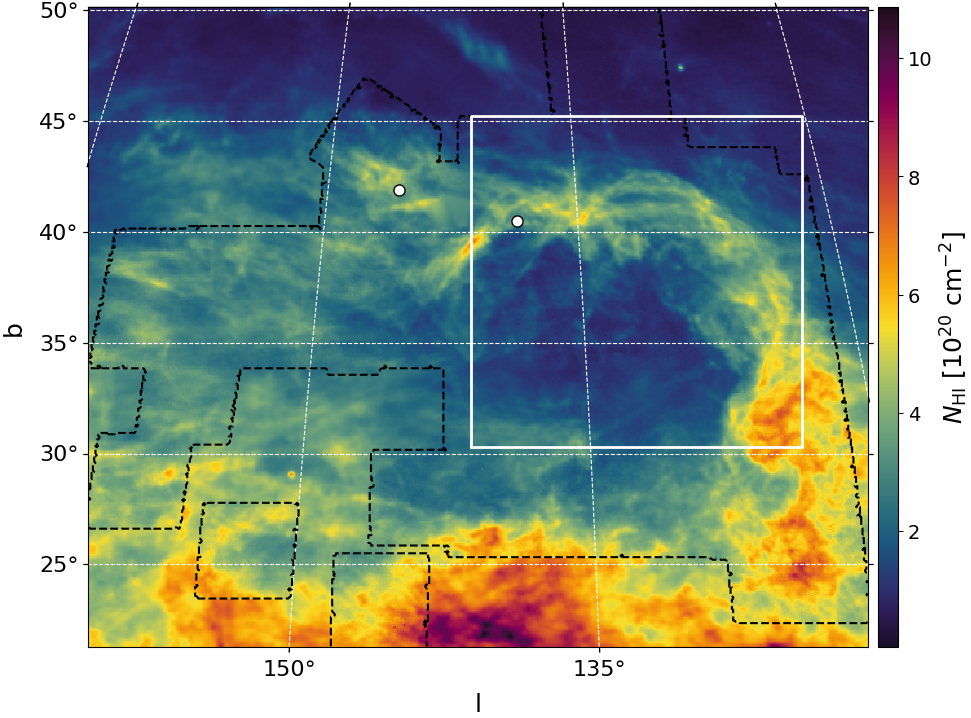}
    \caption{In-painted column density map of the GHIGLS NCPL mosaic integrated over the spectral range $-30.56 < v < 53.17$\,\kms, calculated in the optically thin regime. Outside the black dashed contour are shown data from EBHIS over a similar velocity range ($-29.52 < v < 51.63$\,\kms). 
    The solid white square shows the $256 \times 256$ pixel region on which our analysis is focused. 
    White points represent the positions of the UM (left) and DF (right) absorption sources from the DHIGLS survey. 
    }
    \label{fig:T_b_full_cube_map}
\end{figure*}

Figure~\ref{fig:avgspec} show the mean, median, and standard deviation spectra of the region within the NCPL mosaic that we analysed, the white square in  Figure~\ref{fig:T_b_full_cube_map}. It shows a strong peak of emission near 0\,\kms\ (hereafter, Low Velocity Component or LVC) in which the loop is the most prominent. The secondary broader peak near $-50$\,\kms\ is from what is referred to as the Intermediate Velocity Component (IVC) gas.  
There is also significant emission bridging velocities in between, which does not show the loop and has some characteristics of IVC, but will be distinguished as a Bridge Velocity Component (BVC) in the analysis. In the full NCPL mosaic, there is weak emission in the high velocity range, belonging to the high velocity cloud (HVC) complexes A and C \citep{planck2011-7.12}, but there is none in the region that we analysed. Note that the standard deviation spectrum and mean spectrum have higher contrast relative to each other and to the median spectrum in the LVC compared to BVC and IVC. As we shall see, this is due to the small scale structure in the cold gas that dominates in the LVC.


In this work, we analyse the spectral range of $-30.56 < v < 53.17$\,\kms, which includes adequate emission-free channels on the positive side, and the BVC on the left.\footnote{Specifically, counting from channel 0 in the original cube, this is the 105 channel range starting at channel 200.} 
We specifically want to model the LVC gas and to do that accurately we must include the BVC because its spectral components could overlap with the LVC. On the other hand, the IVC is sufficiently weak and separated from the LVC in this region that it can be ignored. See also Figure~\ref{fig:nh-maps-cnm-scale} in Section~\ref{subsubsec:nh-maps} for validation of this approach.


The \HI spectra in the NCPL are complex, a blend of these different components within which line widths from various thermal phases also differ. This will be discussed in Section~\ref{sec:id-phases}. Spectra are also quite variable spatially.
Figure~\ref{fig:T_b_full_cube_map} shows the total column density map of the GHIGLS NCPL mosaic (inside the black contour) as integrated over the selected velocity range of interest. For completeness, EBHIS data \citep{kerp_2011,winkel_effelsberg-bonn_2016} in the velocity range $-29.52 < v < 51.63$\,\kms were added to cover all latitudes and longitudes shown in Figure~\ref{fig:T_b_full_cube_map}\footnote{Surveyed with the Effelsberg 100-meter radio telescope, EBHIS has a spatial resolution comparable to GHIGLS (10\farcm8 and 9\farcm4 beams, respectively).}.

In this paper, we analyzed a 15\degree\ square region centred near Galactic coordinates $(132\degree, \, +38\degree)$,\footnote{Specifically, counting pixels from (0,0) at the lower left of the full NCPL mosaic, this is the 256 pixel square with lower left corner at coordinate $(295, 154)$.} shown as the white square in Figure~\ref{fig:T_b_full_cube_map}.
This region, which we still refer to as the NCPL, captures the main loop while avoiding some of the extra-galactic emission from the M81/M82 group on the western edge. For the analysis below, pixels contaminated by extra-galactic emission (some even masked in GHIGLS because of corrupted baselines) were in-painted as described in Appendix~\ref{app:inpainting}.

\section{Spectral decomposition} \label{sec:methods}

\subsection{Model} 
\label{sec:model}
We performed a spectral decomposition of NCPL using the publicly available Gaussian decomposition code, 
\ROHSA\footnote{\url{https://github.com/antoinemarchal/ROHSA}}\ \citetalias{marchal_2019}, that was designed to facilitate phase separation using 21-cm data. The algorithm implemented in \ROHSA\ is based on a regularized nonlinear least-square criterion that takes into account the spatial coherence of the emission and the multiphase nature of the gas.

The model $\tilde T_b\big(v, \thetab(\rb)\big)$ used to fit the measured brightness temperature $T_b(v, \rb)$ at a projected velocity $v$ and coordinates $\rb$ is given by:
\begin{equation}
  \tilde T_b\big(v, \thetab(\rb)\big) = \sum_{n=1}^{N} G\big(v, \thetab_n(\rb)\big) \, ,
  \label{eq::model_gauss}
\end{equation}
where each of the $N$ Gaussians, are parameterized by three 2D spatial fields $\thetab_n(\rb)  = \big(\ab_n(\rb) , \mub_n(\rb) , \sigmab_n(\rb) \big)$: amplitude $\ab_{n} \geq \bm{0}$, mean velocity $\mub_{n}$, and standard deviation $\sigmab_{n}$.

\subsection{Optimization of the Gaussian parameters} 
\label{sec:optim}

The parameters $\hat{\thetab}$ are obtained by minimizing the cost function described in Eq.~8 of \citetalias{marchal_2019},
the cost function being a regularized nonlinear least-square criterion containing several distinct ``energy'' terms.
The first term of this cost function is
\begin{equation}
  \label{eq:chi2}
   \frac{1}{2} \, \norm{L\big(v, \thetab\big)}_{\bm{S}}^2 \equiv \frac{1}{2} \, \chi^2(\thetab)
\end{equation}
where $L\big(v, \thetab(\rb)\big)$ is the residual between the model and the data, which is to be compared to the noise $S$ as in the familiar $\chi^2$. To describe the noise properties we adopted the 3D prescription discussed by \cite{boothroyd2011}:
\begin{equation}
  S(v,\rb) = S_{e}(\rb)\, \big(1 + \Trb(\rb)/\Tsys\big) \, ,
  \label{eq:noise}
\end{equation}
where the 2D map of the standard deviation of the noise $S_e(\rb)$ is  calculated from emission-free end channels (in the case of GHIGLS, supplied with the archival data), and $\Tsys$ is the system temperature, typically 20\,K for the GBT L-band observations. 
The original \ROHSA\ code was augmented to work with 3D noise, rather than the standard 2D noise.

In addition, the cost function includes four terms that regularize certain 2D parameter maps of the model. For the first three, a Laplacian filtering is applied on each of the three Gaussian parameter maps $(\ab_n(\rb) , \mub_n(\rb), \sigmab_n(\rb))$ to penalize deviations at high spatial frequencies and ensure a spatially coherent encoding of the signal. The magnitudes of these terms are controlled by the hyper-parameters $(\lambda_{\ab} ,\lambda_{\mub}, \lambda_{\sigmab})$. 

To ensure a solution in which phases are identifiable according to distinct velocity dispersions (phase separation), \citetalias{marchal_2019} introduced a further term in the cost function, $\frac{1}{2}\norm{\bm{\sigma_{n}}-m_{n}}_2^{2}$, where $m_{n}$ is an unknown scalar value to be estimated and $\lambda_\sigma'$ is the corresponding hyper-parameter, again chosen by the user.
This is a global penalty on the variance of each $\sigmab_{n}$ map, minimizing the variance of $\sigmab_{n}$ across the field.  The magnitude of $m_{n}$ found enables an association with a thermal phase of the neutral ISM (i.e., WNM, LNM, or CNM, for large through small values, respectively). 

In this original implementation, the absolute difference of $\sigmab_{n} - m_{n}$ is higher for a broad Gaussian than for a narrow Gaussian, which results in a stronger penalty in the cost function and a narrowing of the distribution of $\sigmab_{n}$ for Gaussians associated with the WNM (see e.g., Figures 16 and 19 in \citetalias{marchal_2019}). To avoid this effect, we use a normalized version,
\begin{equation}
    V(\thetab, \mb) = \frac{1}{2} \, \lambda_\sigma'' \, \sum_{n=1}^N \norm{\frac{\bm{\sigma_n}-m_n}{m_n}}_2^2 \, ,
    \label{eq:V}
\end{equation}
that penalizes the relative difference between $\sigmab_n$ and $m_n$. We denote the hyper-parameter as $\lambda_\sigma''$, rather than $\lambda_\sigma'$ as seen in \citetalias{marchal_2019}, to signal that this normalized version is being used.

In {\tt ROHSA}, starting values of parameters at full resolution are obtained through a multi-resolution process beginning with a fit of the average spectrum of the entire field (\citetalias{marchal_2019}). This procedure aims to facilitate the regularization and to accelerate the search for the initial parameters. The parameters of the Gaussians required to fit this mean spectrum are obtained by a bespoke procedure within \ROHSA\ (\citetalias{marchal_2019}). For later assessment in Section~\ref{subsec:uncertainties}, we altered \ROHSA\ to be able to optimize the fit of the mean spectrum given starting guesses for the parameters, or to simply specify the initial parameters at this top level.

\subsection{Choosing the user parameters}
\label{subsec:user-params}
Application of \ROHSA\ requires choosing values for the number of Gaussians and the four hyper-parameters of the model. These choices need to be revisited for each specific data set \citep[e.g.,][]{marchal_2021b}. To enable a more efficient exploration of these free parameters, we specified that $\big(\lambda_{\ab} ,\lambda_{\mub}, \lambda_{\sigmab} \big)$, which control similar terms in the cost function, be the same.
The decomposition presented in this work was obtained using $\big(N=11, \lambda_{\ab}=\lambda_{\mub}=\lambda_{\sigmab}=100$, and $\lambda''_{\sigmab}=100\big)$.

The choice of the number of Gaussians $N$ is driven by our desire to fully describe the signal while ensuring that the regularization terms in the cost function have the magnitude required to enable a spatially coherent solution and a multi-phase separation.
Given the complexity of the signal coupled with the fact that we ought to describe the LVC and BVCs simultaneously, we explored the parameter space $N=[8-15]$ and made use of the mean contribution to chi-square,

\begin{equation}
    \big<\chi^{2}\big> = \sum_{\rb}\left(\frac{{L\big(v, \thetab(\rb)\big)}}{S(v,\rb)}\right)^2\, /\ 256^{2} \, ,
\label{eq:chi-sq-contrib}
\end{equation}
to find the optimal value. Note, $256^{2}$ is the total number of pixels in our region of focus.

\begin{figure}[!t]
    \centering
    \includegraphics[width=\linewidth]{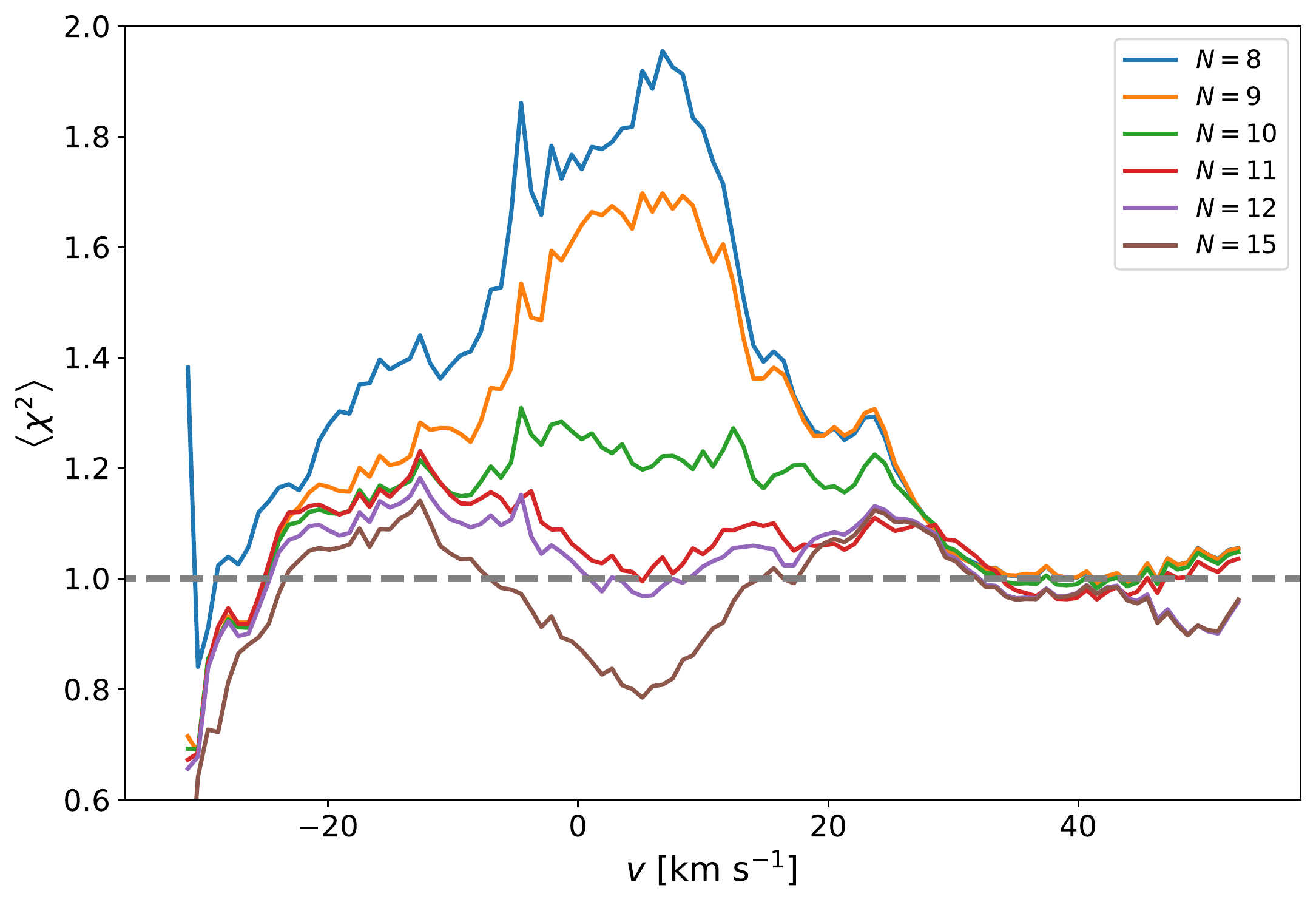}
    \caption{Spectrum of mean contribution to chi-square for Gaussian models fit with different number of Gaussians. Horizontal line indicates $\big<\chi^{2}\big>=1$.}
    \label{fig:chi2-contrib-6-15G} 
\end{figure}

Figure~\ref{fig:chi2-contrib-6-15G} shows the mean contribution to chi-square as function of velocity for increasing values of $N$. The horizontal gray dashed line indicates $\big<\chi^{2}\big>=1$, the expected value for a model close to the data.
As a selection criterion, we chose to keep the model closest to the horizontal line (in the LVC range of interest), $N=11$, without going below it, indicating that the data start to being over-fitted by the model. 

The set of three hyper-parameters controlling the smoothness of the parameter maps was chosen empirically to correlate adjacent pixels on a spatial scale close to the beam of the instrument. In experimenting with this set of parameters, the optimal model was found to have $\lambda_{\ab}=\lambda_{\mub}=\lambda_{\sigmab}=100$. A lower value shows pixelated parameter maps while a higher value alters the native resolution of the observation and prevents the model to being close to the data. Finally, we chose $\lambda''_{\sigmab}=100$ to ensure the phase separation, while not imposing too much penalty on the total cost function.

This specific choice of hyper-parameters does not guarantee that appropriate Gaussians will be found to fit every detail in the cube, due primarily to its complexity. However, in such circumstances it might not matter to the overall goals of the decomposition, even if such a failure is detectable (e.g., in the BVC range), comparing the model to the data.

\subsubsection{Map of $\chi^{2}$}
\label{sec:model-fit}

\begin{figure}[!t]
    \centering
    \includegraphics[width=\linewidth]{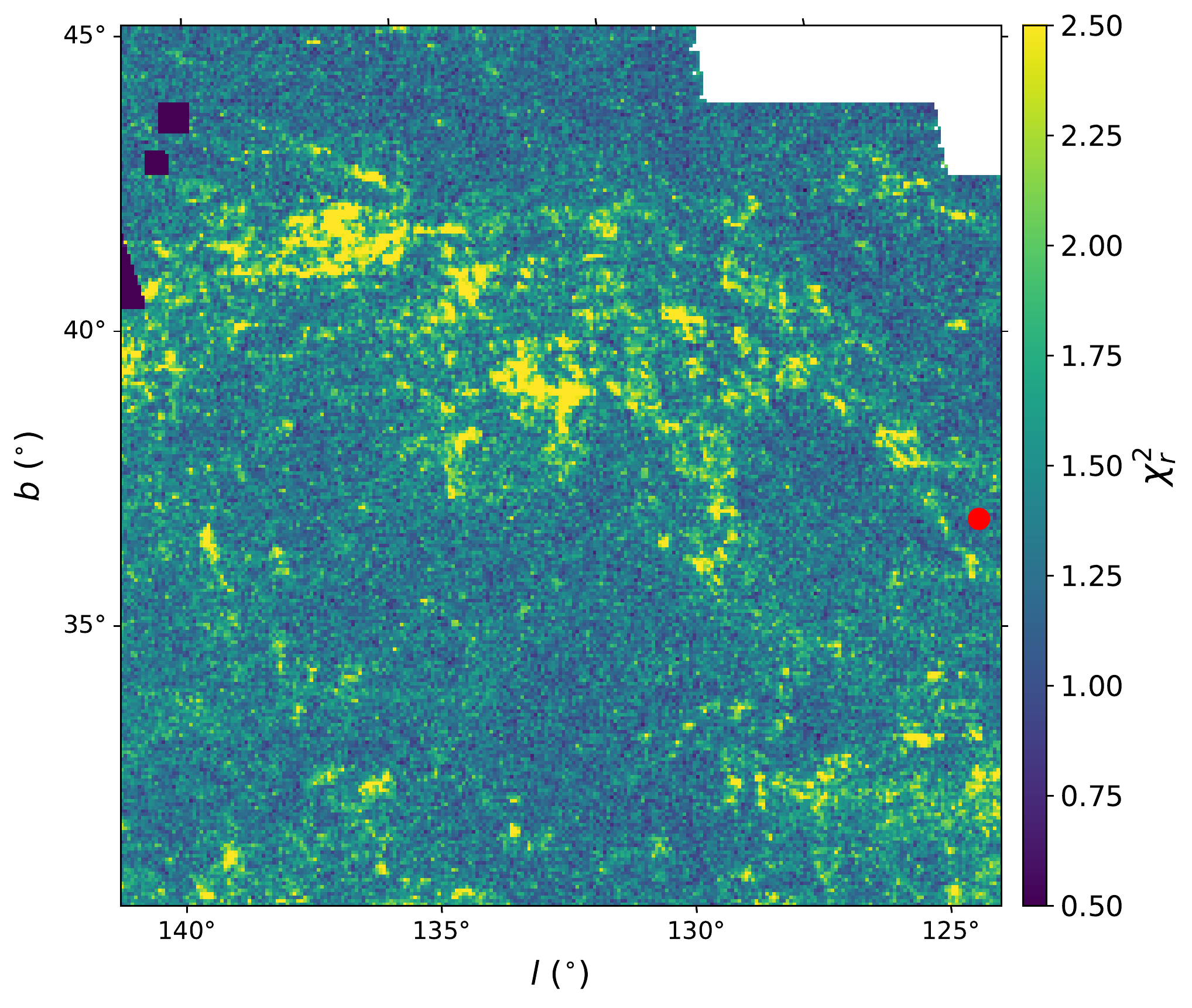}
    \caption{Map of the reduced $\chi^{2}$ obtained with {\tt ROHSA} on the NCPL.
    %
    %
    The location of the spectral decomposition shown in Figure~\ref{fig:random_spectrum} is annotated by the red point.
    }
    \label{fig:mod_rchi2_map}
\end{figure}

Although $\chi^2$ is not the full cost function minimized by \ROHSA, we expect that values for a good fit would be distributed like $\chi^{2}$ for the appropriate number of degrees of freedom. 
We used a reduced and spatially resolved version of Equation~\ref{eq:chi2},
\begin{equation}
  \label{eq:rchi2}
  \chi^2_r(\thetab(\rb)) = \sum_{v} \left(\frac{L\big(v, \thetab(\rb)\big)}{\bm{S}(v,\rb)}\right)^2 / \, k
\end{equation}
where $k = 150 - 3 N_{\rm eff}$ is the number of degrees of freedom, with $150$ the number of velocity channels and $N_{\rm eff}$ the number of Gaussians contributing to the fit. In practice, $N_{\rm eff} < N$ in any pixel because the regularization limits the use of all Gaussians available, especially the term penalizing the variance of each $\sigmab_n$ map. Although a precise map of $N_{\rm eff}$ cannot be predicted, it is possible to approximate it \textit{a posteriori} by counting locally, the number of Gaussians that contribute significantly to the total column density, say at least $1$\%. 

Figure~\ref{fig:mod_rchi2_map} shows the map of $\chi^{2}_{r}$. The dark regions in the upper left are pixels which have been in-painted to avoid modeling extra-galactic emission from the M81/M82 group (see Appendix~\ref{app:inpainting}).
The map is relatively flat, which illustrates that the model fits the data reasonably well. 
The mean, median, and standard deviation are 1.43, 1.37, and 0.39, respectively. In areas where the loop and Spider shape are not seen, the $\chi^{2}_{r}$ is very close to 1.
We found that the areas where $\chi^{2}_{r}$ is highest are due primarily to a poor fit in the BVC range\footnote{Visualization of the residual cube provided a good overview of any deficiencies in the fit, which are most present in the BVC range.}. This does not affect significantly the goodness of fit in the LVC range where the NCPL is located. 

\subsection{A representative solution}
\label{sec:repsol}
\subsubsection{The $\sigma-\mu$ diagram}

\begin{figure*}[!t]
    \centering
    \includegraphics[width=\linewidth]{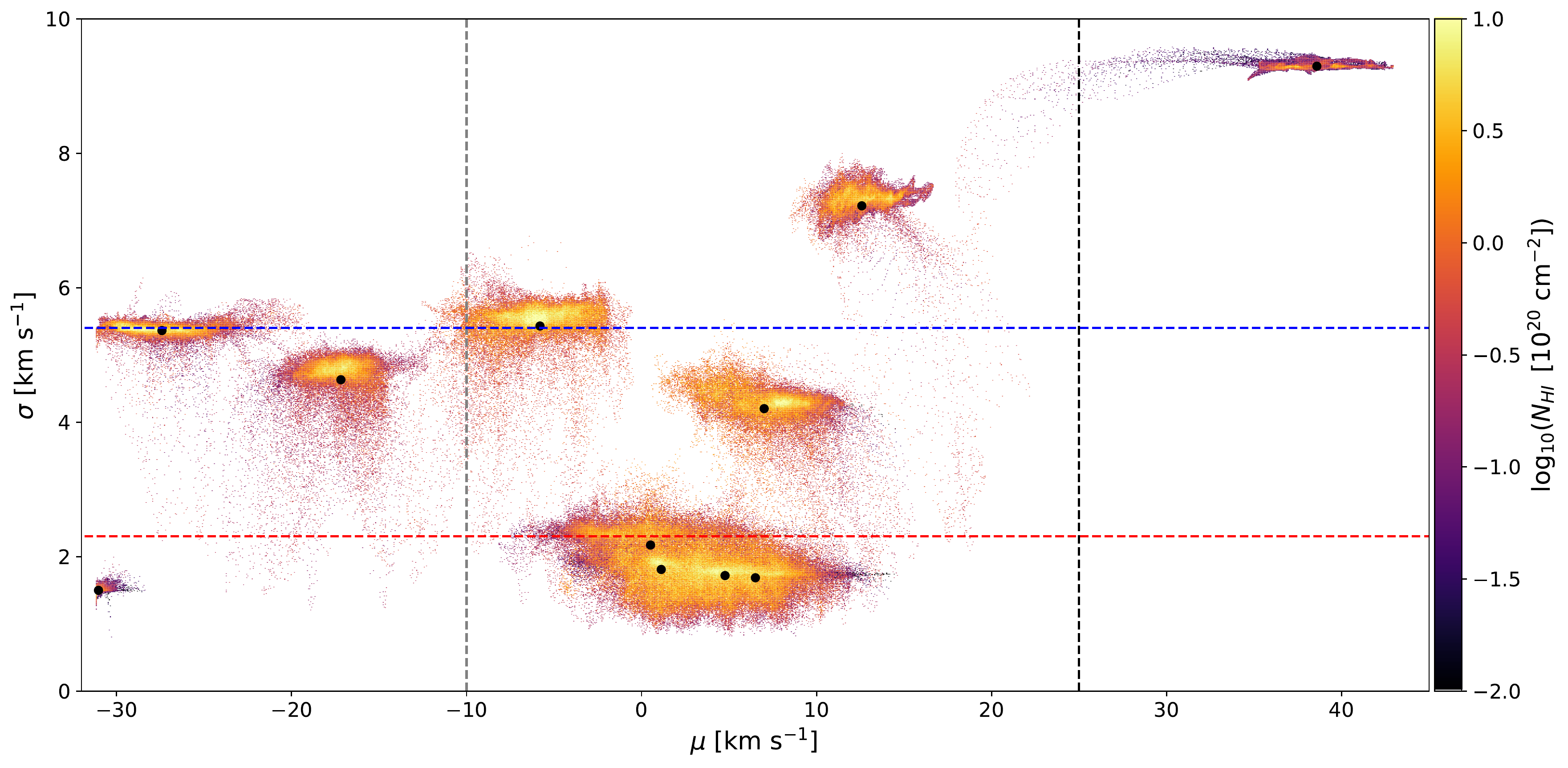}
    \caption{%
      Two-dimensional probability distribution function of \nh\ of the Gaussians in the $\sigma - \mu$ plane.
      Vertical gray line at $\mu=-10\,$\kms\ shows the boundary separating the BVC and LVC.
      Vertical black line at $\mu=25\,$\kms\ represents the outer limit on the LVC range. 
      Horizontal lines at $\sigma=2.3\,$\kms\ (red) and $5.4\,$\kms\ (blue) represent the CNM/LNM and LNM/WNM boundaries (see Section~\ref{sec:id-phases}).
      Black points within the clusters represent the column density-weighted means of each Gaussian cluster, tabulated in Table~\ref{table::mean_var_NCPL}.
      }
    \label{fig:2d_hist}
\end{figure*}

\begin{deluxetable*}{lcccccccccccc}
    \tablecaption{Mean kinematic properties (in \kms) of Gaussians inferred from the NCPL data}
    \label{table::mean_var_NCPL}
    \tablewidth{0pt}
    \tablehead{
    \nocolhead{}  & \colhead{$G_1$} & \colhead{$G_2$} & \colhead{$G_3$} & \colhead{$G_4$} & \colhead{$G_5$} & \colhead{$G_6$} & \colhead{$G_7$} & \colhead{$G_8$} & \colhead{$G_9$} & \colhead{$G_{10}$} & \colhead{$G_{11}$} }
    \startdata
    $\left<\mub_n\right>$ & $-30.2$ & $-26.6$ & $-16.4$ & $-5.0$ & $1.3$ & $1.9$ & $5.6$ & $7.3$ & $7.8$ & $13.4$ & $39.4$ \\ 
    $\left<\sigmab_n\right>$ & $1.5$ & $5.4$ & $4.6$ & $5.4$ & $2.2$ & $1.8$ & $1.7$ & $1.7$ & $4.2$ & $7.3$ & $9.3$ \\ 
    \enddata
\end{deluxetable*}

The corresponding Gaussian parameters for all spatial pixels are summarized in the two dimensional histogram of the column density-weighted $\sigma - \mu$ parameters, seen in Figure~\ref{fig:2d_hist}. There are eleven clusters of points that correspond to the $N$ Gaussians used by {\tt ROHSA} to fit the data. In particular, the separation vertically into clusters of broader and narrower components results from the hyperparameter $\lambda''_{\sigmab}$. Black points show the column density-weighted average of each cluster which are summarized in Table~\ref{table::mean_var_NCPL}. The velocity dispersions of all clusters range from about one to eight \kms. This reveals the multiphase nature of the gas, both in the BVC, and in the LVC range where the NCPL is located.

The corresponding Gaussian parameter maps, sorted by increasing mean velocity, are shown in Appendix~\ref{app:maps}. Rather than present a map of the amplitudes $\ab_n$, we used the analytic integral of the Gaussians to evaluate the column density maps (proportional to $\ab_n \sigmab_n$); these are presented in Figure~\ref{fig:col-dens-mosaic}.
Maps of mean $\mub_n$ and dispersion $\sigmab_n$ are presented in Figures~\ref{fig:vel-mosaic} and~\ref{fig:sigma-mosaic}, respectively.

\subsubsection{LVC and BVC}
\label{sec:id-vel}
Building on this, we refined our LVC/BVC classification based on the Gaussian parameter space. The column density map of each cluster in Figure~\ref{fig:2d_hist} was visualized from Figure~\ref{fig:col-dens-mosaic} to determine if it belongs to the loop. We find that Gaussians $G_4-G_{10}$ shows a loop-like structure and can be classified as LVCs. Their column density weighted velocity centroids are included in the velocity range $-10 < v \, (\kms) < 25$ (see Table~\ref{table::mean_var_NCPL}) denoted by the gray and black vertical dashed lines in Figure~\ref{fig:2d_hist}. Other Gaussians are classified as BVCs. 
As expected from our classification, the resulting total \nh\ maps of the BVC and the LVC (not shown here) do not show any striking correlations.

\begin{figure}[!t]
    \centering
    \includegraphics[width=\linewidth, height=3.005in]{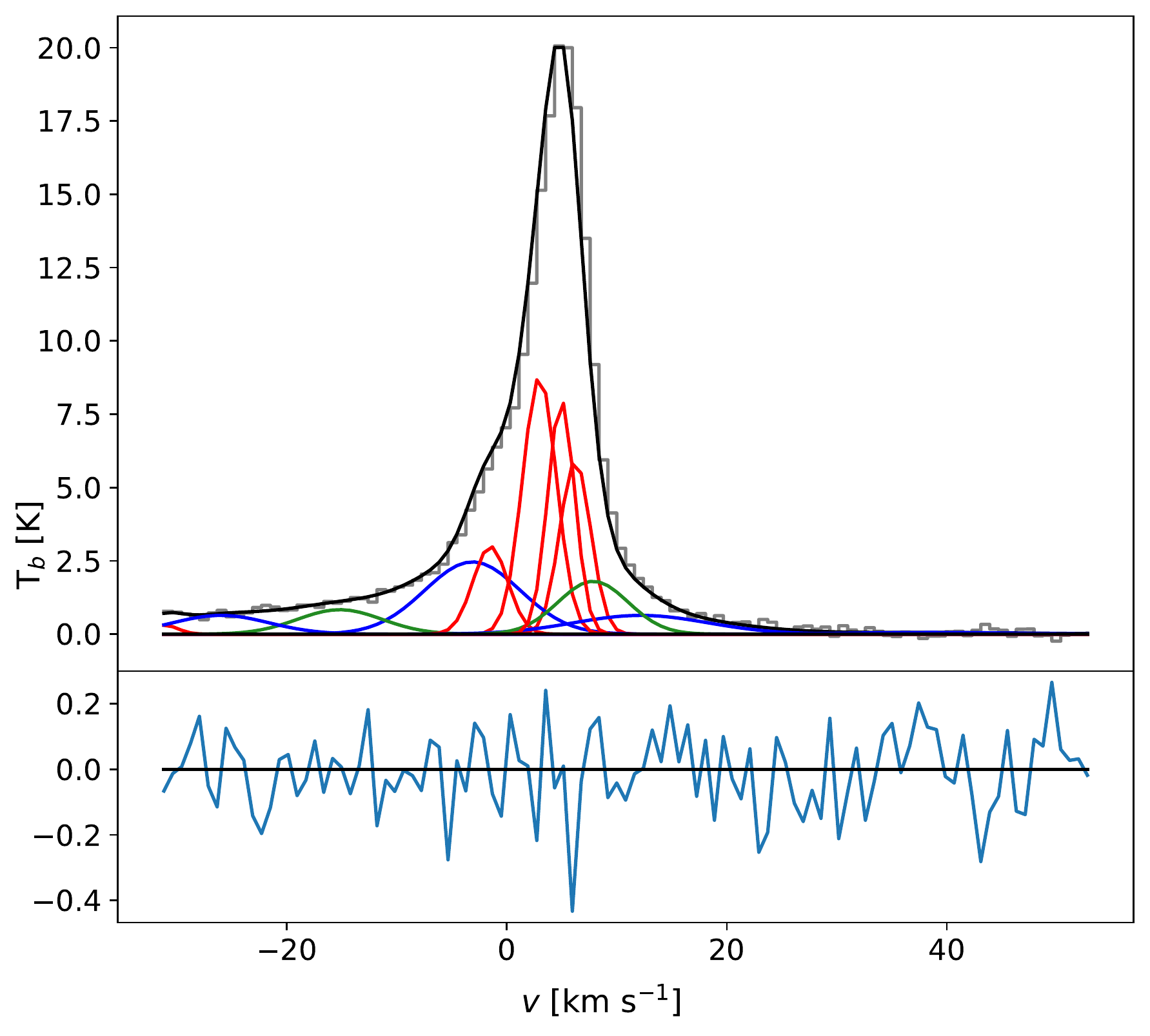} 
    \caption{
    The upper panel is an example of the Gaussian decomposition obtained by \ROHSA\ for the spectrum 
    corresponding to the position annotated by the red point in Fig.~\ref{fig:mod_rchi2_map}.
    Original signal is shown in gray and the total brightness temperature encoded by \ROHSA\ is shown in black. Colored lines represent the individual Gaussian components. Red represents CNM, Green: LNM and Blue: WNM. 
    The lower panel shows the residual spectrum.
    }
    \label{fig:random_spectrum}
\end{figure}

Figure~\ref{fig:random_spectrum} shows an example spectrum (top) where ten out of eleven Gaussians (i.e., $G_1-G_{10}$) are used to fit the data. $G_{11}$ is centered around $\mu=40\,\kms$ and has low amplitude. This particular line of sight, whose location is annotated by the red point in Figure~\ref{fig:mod_rchi2_map}, is representative of both data and decompositions.\footnote{We made use of an original visualization tool wherein we could hover over any pixel in the map and view the corresponding spectrum. We found Figure~\ref{fig:random_spectrum} to be typical of decompositions of the data.} The residual spectrum is shown in the bottom panel. Gaussians $G_4-G_{10}$ encoding the LVC range are, as Figure~\ref{fig:2d_hist} shows, composed of narrow (red), intermediate (green), and wide (blue) components. This specific choice to color code will be discussed in Section~\ref{sec:id-phases}.

\section{Thermal phases in the NCPL}
\label{sec:id-phases}

\subsection{Phase separation}
\label{subsec:grouping}
Focusing on the LVC, mapping out the different thermal phases requires clustering the Gaussian component based on their mean velocity dispersion and mean velocity in Figure~\ref{fig:2d_hist}, and on the morphology of their column density maps in Figure~\ref{fig:col-dens-mosaic}.

Gaussians $G_5-G_8$ are narrow and together form a single large cluster in $\sigma-\mu$ space. Their column density maps display the loop prominently and are structured on small angular scales. We can readily identify the ``legs'' of the Spider. Outside of the loop, however, their column densities are comparatively low. Hereafter, this group of Gaussian clusters describing the cold gas will be simply called CNM.
Gaussians $G_4$ and $G_{10}$ are broad but have very distinct morphologies. While $G_4$ at $\left<\mub_{4}\right>=-5\,\kms$ shows diffuse emission over the entire field including the low density regions below and above the loop, $G_{10}$ at $\left<\mub_{10}\right>=13.4\,\kms$ shows a striking continuous arch. The two warm components described by $G_4$ and $G_{10}$ will be called more memorably WNM$_{\rm D}$ and WNM$_{\rm A}$, respectively, adopting the standard abbreviation WNM for a Warm Neutral Medium ``phase", with subscripts D (``Diffuse") and A (``Arch"). They are kept separate based on their different morphologies and large velocity separation.

Finally, Gaussian $G_9$ has intermediate velocity dispersions, usually associated with lukewarm gas. This will be called LNM. Note that its column density morphology is closer to those of the narrow Gaussians. 

%
On this basis, for subsequent evaluation of the means and uncertainties of column densities we define that the column-density weighted velocity dispersions of CNM, LNM, and WNM Gaussian clusters satisfy $\left<\sigmab_n\right><2.3\,\kms$, $2.3\,<\left<\sigmab_n\right>\,(\kms)<5.4\,$, and $\left<\sigmab_n\right>>5.4\,\kms$, respectively. These boundaries are annotated by the blue and red horizontal lines in Figure~\ref{fig:2d_hist}.
These boundaries translate, for purely thermal broadening, into kinetic temperatures $T_k\sim 520$\,K and $T_k\sim 3400$\,K.
Gaussians shown in Figure~\ref{fig:random_spectrum} are color coded according to this separation (WNM: blue, LNM: green, and CNM: red).
While this operational definition of CNM by grouping of clusters is natural in the context of Gaussian component separation for this specific field, it is neither uniquely definitive nor applicable to others. Note also that part of the broadening can be due to turbulence (Section~\ref{subsubsec:turbulence}), so that the CNM gas would be cooler than 520\,K.

\subsection{Stability analysis and evaluation of uncertainties}
\label{subsec:uncertainties}

\subsubsection{Re-evaluating $N$}
\label{subsubsec:re-evalN}
It is useful at this point to re-evaluate the number of Gaussians needed to describe the emission containing the NCPL. Ensuring that the data are not ``over-fitted" allows us to avoid a possible leakage from the cold phase model to the warm phase model - a broad Gaussians can always be described by a sum of narrow Gaussians. The seven Gaussians previously identified as belonging to the loop were combined to generate a model PPV cube of the LVC gas. Since this model is noise free, we injected a 3D Gaussian random noise using the prescription presented in Equation~\ref{eq:noise} (see Section~\ref{sec:optim}). We applied {\tt ROHSA} on this model cube for $N=[5-8]$ to bracket the original number of Gaussian used to model the LVC gas ($N=7$). The hyper-parameters were kept the same. 

Similarly to Figure~\ref{fig:chi2-contrib-6-15G}, we inspected the mean contribution to chi-square in the velocity range of interest. The decomposition of our model cube with five or six Gaussians prevents us from completely encoding the data. $N=7$ remains closest to the expected value of $\big<\chi^{2}\big>=1$. We retain this number as the optimal choice for describing the multiphase structure of \HI\ gas in the NCPL.

\begin{figure*}[!t]
    \centering
    \includegraphics[width=\linewidth]{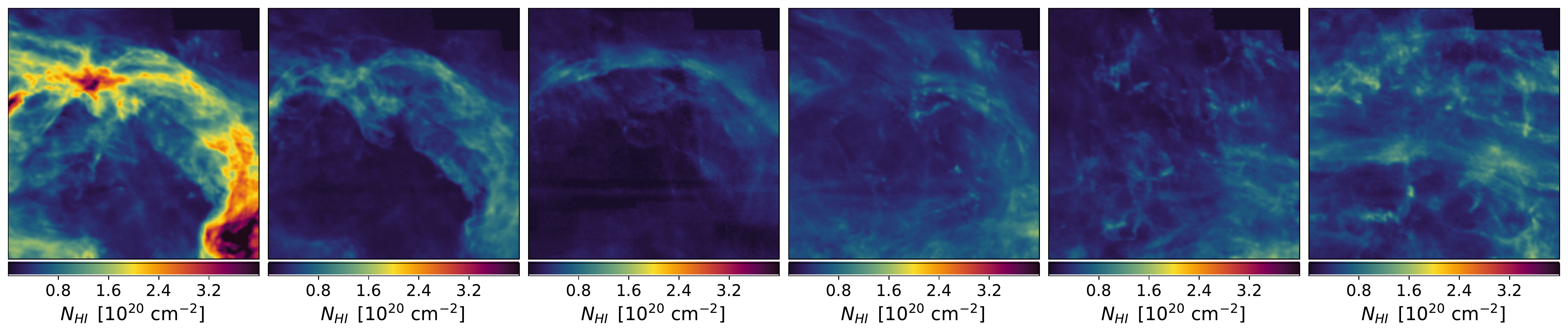}    
    \caption{
    Column density maps of the thermal phases in the LVC, plus maps of the total BVC and IVC, in the NCPL. From left to right, the maps represent the CNM, LNM, WNM$_{\rm A}$, WNM$_{\rm D}$ (all of the LVC), BVC and IVC. Maps have the same dynamical range (color bar) to highlight the dominance of the CNM.
    }
    \label{fig:nh-maps-cnm-scale}
\end{figure*}

\subsubsection{Uncertainties}
\label{subsubsec:uncertain}

Although we have determined the optimal number of Gaussians needed to describe the NPCL data in the LVC range, it remains that their position in the parameter space can vary due to the non-linear nature of the model. There are several possible combinations/arrangements of the seven Gaussians that can be used to obtain a solution that satisfies our quality criteria (see Section~\ref{subsec:user-params}).

To explore the degeneracy of the solution, we have repeated the Gaussian decomposition of the model cube using three series of runs. The first two series, composed of 50 runs each, follow the methodology developed by \citet[][see their section~4.1.1]{marchal_2021b}.
The first series explores how the outcome of the decomposition is influenced by the injection of 50 different instance of the noise. The hyper-parameters were kept the same. 
The second series explores the impact of the hyper-parameters on the solution. It entails generating $50$ runs using random perturbations of the four \ROHSA\ hyper-parameters in a $\pm10$\% interval around the original values. Here, the injected 3D Gaussian random was kept the same.

The third series is original to this work. It explores the sensitivity to how \ROHSA\ is initialized (i.e., choice of the initial parameters at the top level of the multi-resolution process, or descent; see Section~\ref{sec:optim}).
It entails generating 50 runs where the seven Gaussians needed to initialized the descent are randomly selected from the original $\sigma$ -- $\mu$ diagram shown in Figure~\ref{fig:2d_hist}. Specifically, we selected pixels within the following constraints: $5\,$\kms\ in $\mu$ and $1\,$\kms\ in $\sigma$ relative to the column density weighted means of each Gaussian cluster. This ensures that these Gaussians provide a good average representation of the data. Here, the hyper-parameters were kept the same as the original run.

For each of the 50-run series, the outcomes were examined in $\sigma$ -- $\mu$ space, showing that the clusters observed in Figure~\ref{fig:2d_hist} are quite stable, including the need for the intermediate LNM component.

For each run, we performed a phase separation by grouping the seven Gaussians into the four components CNM, LNM, WNM$_{\rm D}$, and WNM$_{\rm A}$ based on the mean velocity dispersion limits established in Section~\ref{subsec:grouping}. Then for each series, maps of the mean column density and its standard deviation were generated for the four components. Intercomparison of these results between series revealed that they were compatible.

Therefore, all 50 runs of each series were combined to calculate maps for two phase-separated quantities of interest discussed in the following section: column density and mass fraction, and their standard deviations.
Finally, for each quantity the contributions from the three series were summed in quadrature to yield the total uncertainty (the contribution from the third series was generally slightly higher than from the first and second).

\subsection{Column density and mass fraction maps}
\label{subsubsec:nh-maps}

\begin{figure*}[!t]
    \centering
    \includegraphics[width=\linewidth]{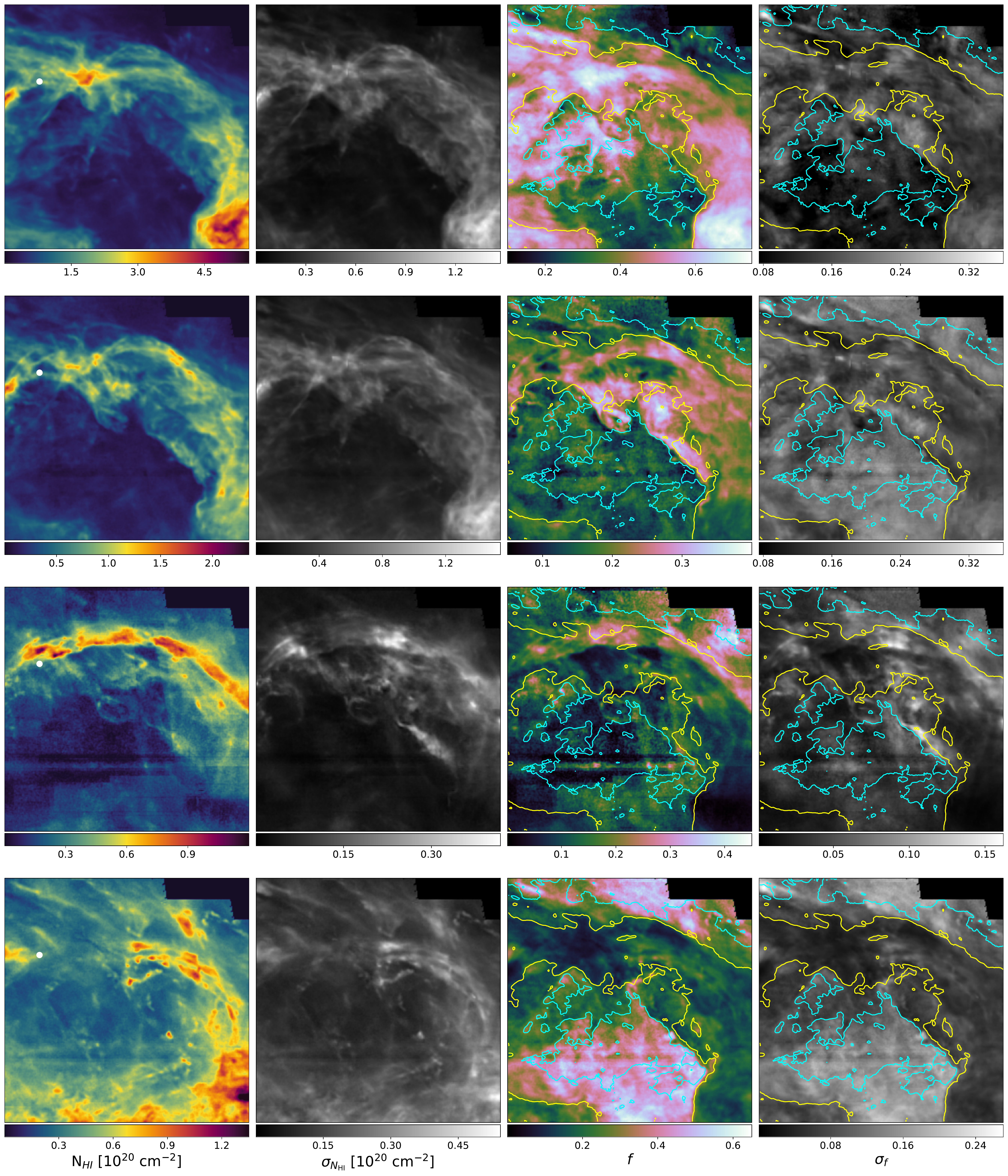}
    \caption{
    Column density and mass fraction maps (first and third column) with their associated standard deviation maps (second and fourth column) of the thermal phases in the NCPL, obtained using the means and standard deviations of the three methods described in Section~\ref{subsubsec:uncertain}. Rows from top to bottom show the CNM, LNM, WNM$_{\rm A}$, and WNM$_{\rm D}$, respectively. White points in the first column represent the position of the DF absorption source from the DHIGLS survey. Cyan and yellow lines represent contours of the total LVC column density with levels of $1\times10^{20}\,$ and $2\times10^{20}\,$cm$^{-2}$ respectively.
    }
    \label{fig:nh-std-maps}
\end{figure*}

Figure~\ref{fig:nh-maps-cnm-scale} shows the \nh\ maps in column one of Figure~\ref{fig:nh-std-maps}, as well as the \nh\ maps of the BVC and IVC, but all with the same dynamical range (color bar) as in the CNM map. This shows just how abundant the cold gas in the LVC is, and further verifies the large mass fraction of the CNM compared to the other phases and velocity components.

\begin{deluxetable}{lccccccc}
    \tablecaption{Mean column density and mass fraction properties and their uncertainties of phases inferred from NCPL data ``on loop" and ``off loop" using a mask with level $2\times10^{20}\,$cm$^{-2}$ from the total column density of the LVC model cube}
    \label{table::mean_prop}
    \tablewidth{\linewidth}
    \tablehead{
    \nocolhead{}  & \colhead{$\langle \nh \rangle$} & \colhead{$\langle \sigma_{\nh} \rangle$} & \colhead{$\langle f \rangle$} & \colhead{$\langle \sigma_{f} \rangle$} \\
    \nocolhead{}  & 10$^{20}$ cm$^{-2}$ & 10$^{20}$ cm$^{-2}$ & &  
    }
    \startdata
    \textbf{On loop} \\
    CNM & 1.71 & 0.50 & 0.50 & 0.15 \\ 
    LNM & 0.71 & 0.50 & 0.21 & 0.15 \\ 
    WNM$_{\rm A}$ & 0.12 &  0.08 & 0.11 & 0.04 \\
    WNM$_{\rm D}$ & 0.25 & 0.12 & 0.16 & 0.08 \\
    \hline
    \textbf{Off loop} \\
    CNM & 0.42 & 0.14 & 0.36 & 0.12 \\ 
    LNM & 0.19 & 0.16 & 0.16 & 0.14 \\ 
    WNM$_{\rm A}$ & 0.15 & 0.05 & 0.13 & 0.04 \\
    WNM$_{\rm D}$ & 0.34 & 0.16 & 0.32 & 0.15 \\
    \enddata
\end{deluxetable}

Figure~\ref{fig:nh-std-maps} shows column density and mass fraction maps (first and third column) with their associated standard deviation maps (second and fourth column) of the thermal phases in the NCPL.
Rows from top to bottom show the CNM, LNM, WNM$_{\rm A}$, and WNM$_{\rm D}$, respectively. To facilitate the localization of the loop in the mass fraction maps, we overlaid contours of the total LVC column density using levels of $1\times10^{20}\,$ (cyan) and $2\times10^{20}\,$cm$^{-2}$ (yellow). Note that the main attributes of the Spider, like its body and ``legs" are all within the cyan and yellow contours. 
The average properties (column density and mass fraction) for each component are tabulated in Table~\ref{table::mean_prop} where we distinguished ``on loop" and ``off loop" using the $2\times10^{20}\,$cm$^{-2}$ level shown by the yellow contour. 

%
The CNM component (first row) displays the loop prominently. We can identify the bright body of the Spider as well as its individual ``legs". Others areas of highest column densities include Polaris located in the south-west part of the field and a body located at the eastern edge of the field, toward Ursa Major. The CNM mass fraction ``on loop" and ``off loop" are $0.50\pm0.15$ and $0.36\pm0.12$, respectively. Locally, in the body of the Spider and the southern part of Polaris, the CNM mass fraction range up to about $0.72$.

The LNM (second row) also shows the loop prominently but with a lower column density contrast. The body of the Spider is less pronounced and shows small scale structure around it. We can identify the bright body toward Ursa Major in this map as well. As with the body of the Spider, Polaris appears less bright. The overall mass fraction in the LNM is lower than in the CNM with an average ``on loop" and ``off loop" of $0.21\pm0.15$ and $0.16\pm0.14$, respectively. Note that standard deviation maps of LNM and CNM are highly correlated, highlighting the degeneracy of the solution. These numbers should be considered carefully and their combined mass fraction considered a more robust estimate of the amount of cold and lukewarm gas mixed together. 

\begin{figure}[!t]
    \centering
    \includegraphics[width=\linewidth]{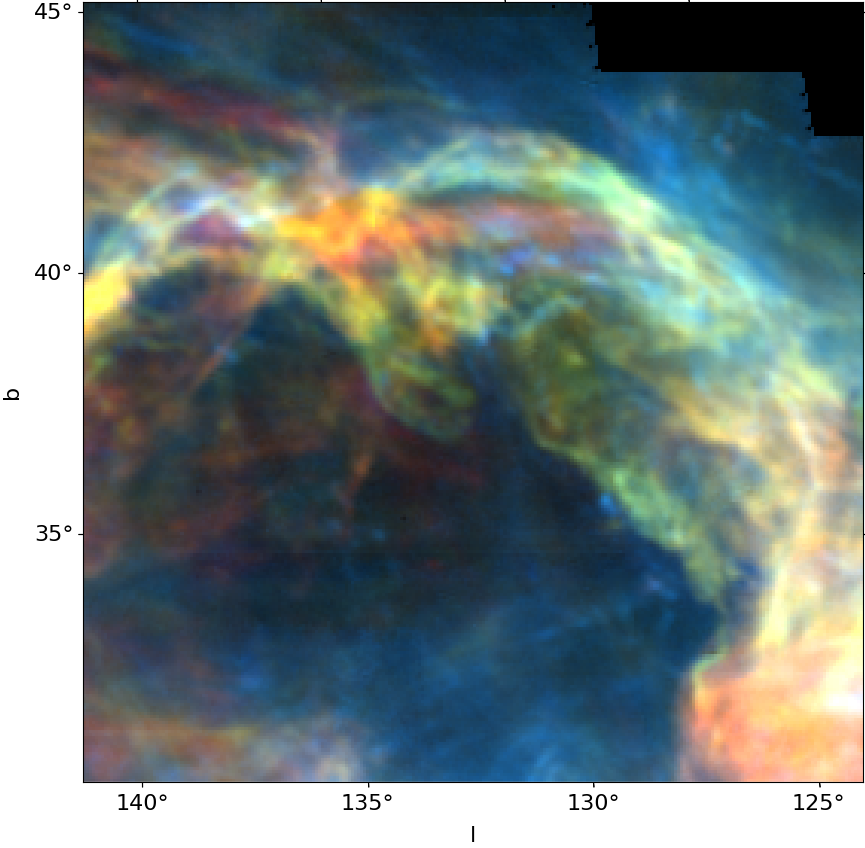}    
    \caption{
    Three-phase view of the NCPL from column density maps of CNM (red), LNM (green), and WNM$_{\rm D}$+WNM$_{\rm A}$ (blue). Maps were scaled with natural weightings to highlight their interrelationship and relative column densities.
    }
    \label{fig:NCPL_rgb}
\end{figure}

As presented previously, WNM$_{\rm A}$ shows a continuous arch. It coincides spatially with the top of the loop traced by the cold and lukewarm gas from the center of the field to the west. Toward the east, ``legs" come out of the body of the Spider and no longer follow the warm arch. This can be visually appreciated in the WNM$_{\rm A}$ mass fraction map where contours of the loop are displayed, and in the three-phase RGB view shown in Figure~\ref{fig:NCPL_rgb} where phases CNM, LNM, and WNM$_{\rm A}$+WNM$_{\rm D}$ are color coded by red, green and blue, respectively. Mass fractions ``on loop" and ``off loop" are similar, with values of $0.11\pm0.04$ and $0.13\pm0.04$, respectively.

WNM$_{\rm D}$ (fourth row), is much more diffuse on large scales but has bright features on small scales at various locations in the field that sometimes do not coincide with the loop. The fact that WNM$_{\rm D}$ overlaps with the BVC range, which shows similar structures at small scales, may be a clue that these clouds physically coincide with IVCs located further along the line of sight but appear at lower velocities in projection. As for the CNM, mass fractions ``on loop" and ``off loop" are significantly offset, with values of $0.16\pm0.08$ and $0.32\pm0.15$, respectively. In the southern part of the field, which coincides with the low total column density region of the NCPL (i.e., in the inner part of the loop), the mass fraction of warm gas range up to 0.61.

\subsubsection{Dependence on phase separation boundaries} 

We compared our boundaries for the LVC in the NCPL field to those found by \citet{kal-haud-2018} over the full sky from their Gaussian decompositions. In their figure 5, the boundary between CNM and LNM is about $5.62\,$\kms\ FWHM ($\sigma = 2.39\,$\kms) and between LNM and WNM about $15.84\,$\kms\ FWHM ($\sigma = 6.73\,$\kms). These are (coincidentally) quite comparable to the boundaries $\sigma=2.3\,$\kms\ (CNM/LNM) and $\sigma=5.4\,$\kms\ (LNM/WNM) discussed above. The boundaries vary from field to field given their unique dynamical histories and we are confident in the particular choices we have made.

Looking at the LVC clusters with low $\sigma$ in Figure~\ref{fig:2d_hist}, if we increased the red boundary anywhere up to 4\,\kms\ there would no difference in what we call CNM. 
On the other hand, if we lowered the blue boundary, cluster $G_4$ could become LNM.  This would substantially increase the total LNM, adding gas at a distinctly different velocity.

Moving the LNM/WNM boundary as low 4.5\,\kms\ would make no difference, but if we increased it, then WNM$_{\rm D}$
would be re-classified as LNM. Given its very different morphology, we would call it LNM$_{\rm D}$ and not just add it to the other LNM cluster. 
However, the CNM would still dominate in terms of mass fraction. Indeed, having LNM dominate would seem very unusual.

Moving the boundaries would also impact the standard deviation maps in columns 2 and 4 of Figure~\ref{fig:nh-std-maps}. Most significantly, raising the CNM/LNM boundary 
would actually decrease our estimate of the uncertainty of the CNM and LNM column densities and their mass fractions, because there would be fewer Gaussians in the 150 runs where the slight change of the dispersion would be enough to have it cross the boundary and be reclassified as the other phase.
%

\subsection{Anti-correlations between phases} \label{subsec:corr}

\begin{figure}[!t]
    \centering
    \includegraphics[width=\linewidth]{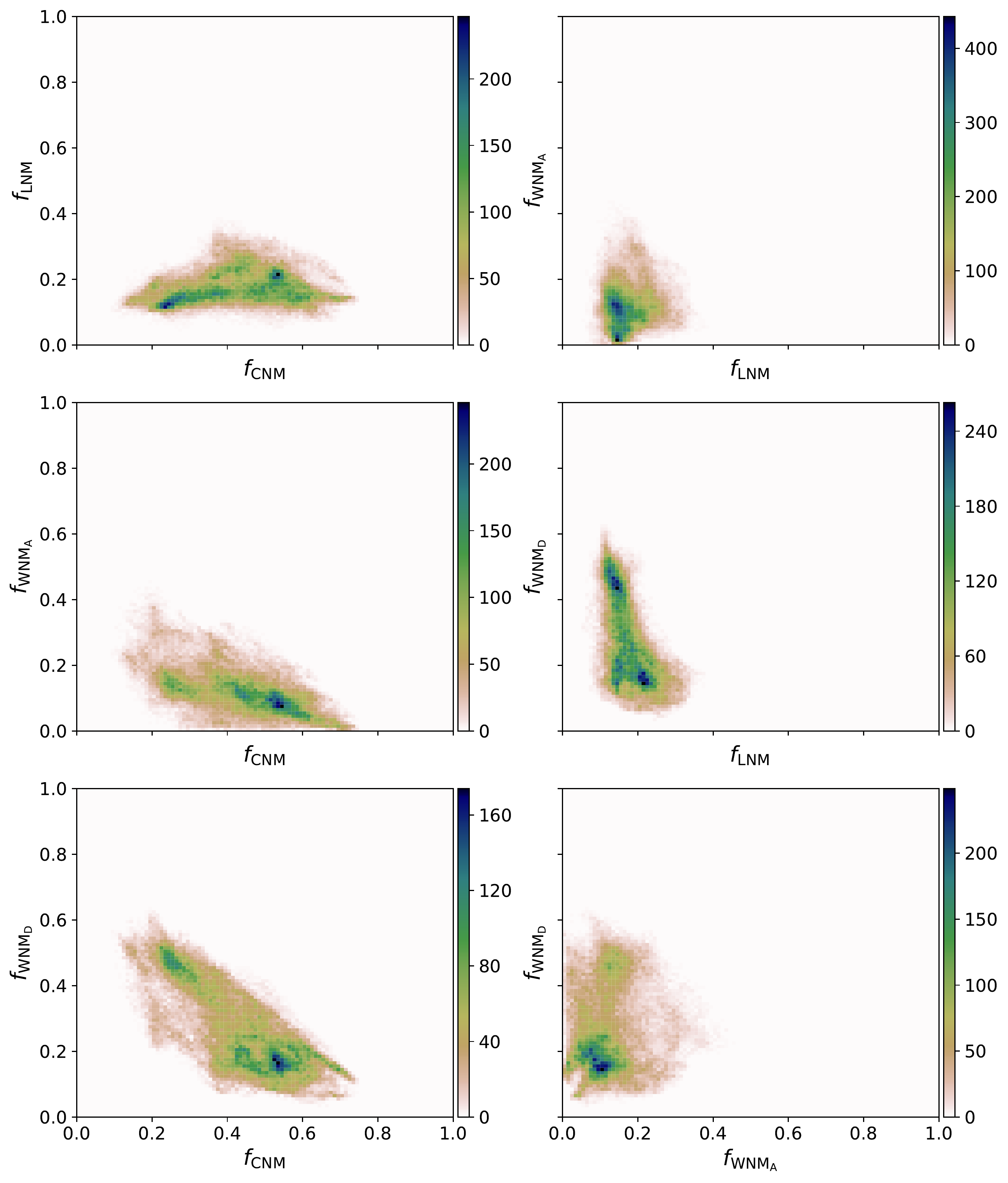}    
    \caption{2D distribution functions, showing how the various phase mass fractions are related to each other. Left column shows the relationship between LNM (top), WNM$_{\rm A}$ (middle) and WNM$_{\rm D}$ (bottom) and the CNM. First two rows of the right column show the relationship between the WNM components (WNM$_{\rm A}$: top, WNM$_{\rm D}$: middle) and the LNM. Bottom right panel shows the relationship between the two WNM components.
    }
    \label{fig:correlations}
\end{figure}

If the WNM mass fraction is low, as found by \citet{kal-haud-2018} in their figure 18 right, then the LNM mass fraction and the CNM mass fraction must anti-correlate, by construction (the fractions need to add to 1). In our field covering the NCPL loop and its surroundings, the situation is much more complex, as shown in Figure~\ref{fig:correlations}.
Both $f_{\rm LNM}$ and $f_{\rm WNM_A}$ are relatively small and so $f_{\rm WNM_D}$ and $f_{\rm CNM}$ show the most obvious anti-correlation (lower left panel). 
This can be seen in the morphology of the maps of $f_{\rm WNM_D}$ and $f_{\rm CNM}$ in column 3 of Figure~\ref{fig:nh-std-maps}, using the column density contours for reference. The upper left endpoint of the anti-correlation corresponds to the lower central area of the map and the area above the loop, where WNM$_{\rm D}$ dominates, while the lower right endpoint corresponds to the loop, where CNM dominates.
Along the loop, in places where there is little WNM, CNM and LNM can be seen to anti-correlate and this is reflected in the upper left panel of Figure~\ref{fig:correlations}. But elsewhere, there are areas where all phases contribute along the line of sight.

\section{Further evidence of cold gas from absorption lines in the NCPL}\label{sec:absorption1}

\subsection{Absorption against 4C +71.09}
\label{subsec:4c+71.09}

\begin{deluxetable*}{lccccccccccccc}
\tabletypesize{\footnotesize}
\tablecolumns{12} 
\tablewidth{0pt}
\tablecaption{Parameters\tablenotemark{a} of select absorption features, interpolated emission, and derived $\Trs$ from DHIGLS and GHIGLS}
\label{table:abs}
\tablehead{
\colhead{Region} & \colhead{NVSS Source} & \colhead{$\Trc$} & \colhead{$\mu_{\mathrm n}$} & \colhead{$\sigma_{\mathrm n}$} & \colhead{$\Trn$} & \colhead{$\mu_{\mathrm b}$} & \colhead{$\sigma_{\mathrm b}$} & \colhead{$\Trb$} & \colhead{$\Trs$} & \colhead{$\sigma_{\rm th}$} & \colhead{$\sigma_{\rm nt}$} & \colhead{$f_{\rm CNM}$} & \colhead{$M_t$}
}
\startdata
\DFG & J$101132+712440$ & 380 & $5.63$ & $1.30$ & $-75$ & $5.89$ & $1.58$ & 19.11 & 77 & 0.80 & 1.03 & 0.42 & 2.6 \\ 
\DFG & J$101132+712440$ & 380 & $5.63$ & $1.30$ & $-75$ & $6.00$ & $1.50$ & 15.20 & 64 & 0.73 & 1.08 & 0.32 & 3.0 \\
NCPL & J$101132+712440$ & 380 & $5.63$ & $1.30$ & $-75$ & $5.33$ & $2.02$ & 17.56 & 72.1 & 0.77 & 1.05 & 0.54 & 2.8 \\
\hline
\UMG & J$094912+661459$ & 444 & $2.43$  & $1.52$ & $-109$ & $2.67$ & $1.77$ & 22.41 & 76.3 & 0.80 & 1.31  & 0.51 & 3.3 \\
\UMG & J$094912+661459$ & 444 & $2.43$  & $1.52$ & $-109$ & $2.66$ & $1.99$ & 22.73 & 72.2 & 0.80 & 1.30 & 0.55 & 3.4 \\
\enddata
\tablenotetext{a}{Velocities in \kms\ and temperatures in K.
}
\end{deluxetable*}

To further investigate the presence of CNM in the NCPL, it is useful to turn to the DF field from the DHIGLS \HI\ survey\footnote{DRAO \HI Intermediate Galactic Latitude Survey: \url{https://www.cita.utoronto.ca/DHIGLS/}} \citepalias{blagrave_dhigls:_2017} with the Synthesis Telescope (ST) at the Dominion Radio Astrophysical Observatory. DF is embedded in the NCPL field of GHIGLS and provides a spatial resolution of about 1', allowing the detection of radio point sources that have a high continuum brightness temperature, $\Trc.$\footnote{$\Trc$ depends on the cataloged source flux density and inversely as the square of the beam size; see equation 8 in \citet{blagrave_dhigls:_2017}.} The velocity resolution and channel spacing of the spectrometer were 1.32\,\kms and $\Delta v=0.824$\,\kms, respectively.

\subsubsection{Spin temperature}
In their paper, \citetalias{blagrave_dhigls:_2017} reported the detection of CNM gas in DF against the background radio galaxy ${\rm 4C+71.09}$ \citep[NVSS\footnote{NRAO VLA Sky Survey:\ \url{https://www.cv.nrao.edu/nvss/}} ${\rm J101132+712440}$,][]{NVSS-1998}, located within the loop as shown by the white points in Figures~\ref{fig:T_b_full_cube_map} and \ref{fig:nh-std-maps}. They found a single cold component with spin temperature $\Trs = 84$ K from  
\begin{equation}
    \Trs = \Trc \, \Trb / (\Trb \, - \, \Trn) \, ,
    \label{eq:Ts}
\end{equation}
where $\Trb$ is the brightness temperature of the cold gas from the emission spectrum interpolated at the position of the source, and $\Trn$ is the net temperature of the absorption spectrum. These numbers were taken visually from the original DF data (see Section~8.3 in \citetalias{blagrave_dhigls:_2017} for further details). In this work, we refine our $\Trs$ estimate by modeling the emission-absorption pair with Gaussians to allow for a better estimation of the cold component amplitude.

The absorption spectrum at the position of the source is well described by a single Gaussian component superimposed on a quadratic polynomial to compensate the baseline residual present in the data.
Parameters $\mu_{\mathrm n}$, $\sigma_{\mathrm n}$, and $\Trn$ of the Gaussian are tabulated in Table~\ref{table:abs}.

Modeling the emission spectrum requires approximating the total brightness temperature emitted by the gas that would be observed in the absence of the background radio source. This is usually achieved by linear interpolation of the emission spectra in the area surrounding the source \citep[e.g., ][]{heiles_millennium_2003-1,murray2015,murray_2018,syed_2021}. 
Specifically, we calculated the average spectrum taken in an annulus centered on the source with an inner radius $r_{\rm in} = 1\farcm2$ and an outer radius $r_{\rm out} = 3'$ (i.e., 4 and 10 pixels of size 18\arcsec, respectively). 

\begin{figure}[!t]
    \centering
    \includegraphics[width=\linewidth]{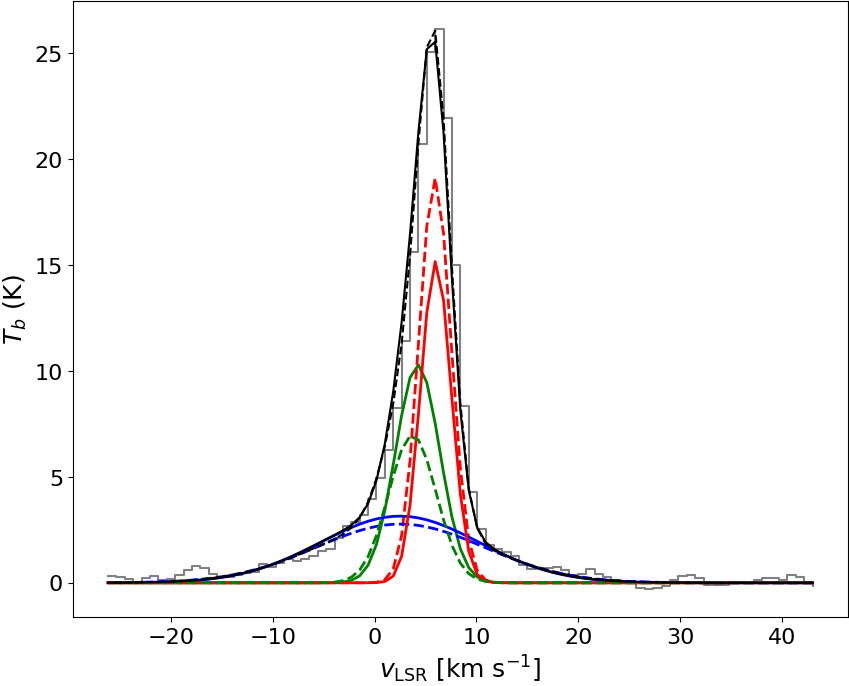}
    \caption{Gaussian decomposition of the averaged emission spectrum taken in a $\sim2'$ annulus centered on the source. The original signal is shown in gray and the total signal encoded by the Gaussians in black. The three individual Gaussians (solid lines) are color coded by phase, same as Figure~\ref{fig:random_spectrum}. 
    Gaussian decomposition by {\tt ROHSA} from interpolation of the parameter maps in the same annulus is shown with dashed lines and the same color code.
    Parameters of the narrowest Gaussians are tabulated in Table~\ref{table:abs}.
    }
    \label{fig:DF_abs_source_interp_ROHSA_native_res}
\end{figure}

The resulting spectrum is well fitted by a sum of three Gaussians (one for each phase) as shown by the solid lines in Figure~\ref{fig:DF_abs_source_interp_ROHSA_native_res}.
Parameters $\mu_{\mathrm b}$, $\sigma_{\mathrm b}$, and $\Trb$ of the narrowest Gaussian are tabulated in Table~\ref{table:abs} (first row). 
Using the numbers from Table~\ref{table:abs}, Equation~\ref{eq:Ts} evaluates to $\Trs = 77$\,K. This is lower than found by \citetalias{blagrave_dhigls:_2017} and is explained by the presence of an intermediate Gaussian, possibly modeling unstable gas, that lowers the amplitude of the cold component. We find a CNM mass fraction, $f_{\rm CNM}=0.43$ (see Table~\ref{table:abs}).

Alternatively, we performed a Gaussian fit of a $64\times64$ pixel grid centered on the source with \ROHSA. In that case, three Gaussians are needed to fully encode the signal as well, as shown by the dashed lines in Figure~\ref{fig:DF_abs_source_interp_ROHSA_native_res}). Parameters $\mu_{\mathrm b}$, $\sigma_{\mathrm b}$, and $\Trb$ of the narrowest Gaussian at the position of the source are interpolated from the {\tt ROHSA} parameter maps in the same annulus and are also tabulated in Table~\ref{table:abs} (second row). Equation~\ref{eq:Ts} evaluates to $\Trs=64$\,K. Here, we find $f_{\rm CNM}=0.33$ (see Table~\ref{table:abs}).

In these two alternative decompositions the spectral data being fit on the $64\times64$ pixels are identical, and so this comparison demonstrates the degeneracy of the solutions that can arise when fitting a non-linear model like a sum of Gaussians and the sensitivity of the values of derived parameters.

\subsubsection{Thermal and non-thermal broadening}
\label{subsubsec:turbulence}

In the CNM, it is believed that collisional excitation alone allows $\Trs \simeq T_{\rm k}$ \citep{murray2015}, where $T_{\rm k}$ is the kinetic temperature. It follows that the thermal broadening of the cold component
\begin{equation}
    \sigma_{\rm th} = \sqrt{T_{\rm k}\,k_{\rm B}\,/\,{m_{\rm H}}}\, \simeq \sqrt{\Trs\,/\,{121}} \, ,
    \label{eq:sig_th}
\end{equation}
where $m_{\rm H}$ is the mass of hydrogen, $k_{B}$ is the Boltzmann constant, and the numerical factor is in units of K\,km$^{-2}$\,s$^{2}$. For both of the above methods, Equation~\ref{eq:sig_th} evaluates to a smaller dispersion than that of the total broadening of the absorption line, $\sigma_{\mathrm n}$, that measures the total velocity dispersion of gas along the line of sight, including systematic motions (e.g., velocity gradients), instrumental broadening\footnote{From the channel spacing of the spectrometer, $\Delta v$, we found an instrumental line broadening of about $0.36\,$\kms, by fitting a Gaussian to a zeroed spectrum, with a single channel set to $1\,$K.}, and/or turbulent motions from the energy cascade.

The total broadening of the line is often modeled as a quadrature sum of a thermal component and a non-thermal component, 
\begin{equation}
    \sigma_{\rm n}^2 = \sigma_{\rm th}^2 + \sigma_{\rm nt}^2 \,,
    \label{eq:quadratic-disp}
\end{equation}
\citep[e.g.,][]{heiles03,marchal_2021a, marchal_2021b,syed_2021}. Rearranging Equation~\ref{eq:quadratic-disp} to use the entries in the table leads to $\sigma_{\rm nt} \approx 1 \, \kms$, as entered in Table~\ref{table:abs}. 

The turbulent Mach number is given by
\begin{equation}
    M_t^2 = 3\, \sigma_{\rm nt}^2 / C^2_{\rm iso} = 4.2\, \sigma_{\rm nt}^2/\sigma_{\rm th}^2 \,,
    \label{eq:mach}
\end{equation}
where $C_{\rm iso} = \sqrt{\sigma_{\rm th}/\mu}$ is the isothermal sound speed, appropriate in the CNM.\footnote{The mean molecular weight is $\mu m_{\rm H}$ with $\mu = 1.4$. Combined with Equation \ref{eq:quadratic-disp}, Equation~\ref{eq:mach} is equivalent to equation 17 of \citet{heiles03}.}  Values of $M_t$ are calculated from the dispersions already in the table.

We find $M_t\approx3$. Although the NCPL is a specific environment, this value is close to the peak of the histogram from all 21-Sponge survey sources analysed by \cite{murray2015} (see their figure~5) and consistent with the earlier findings of \citet{heiles03}, regardless of the environment to which they belong. This could be a coincidence, given that the histograms have a considerable spread.

Comparison of the parameters for absorption and emission reveals that $\mu_{\rm b}$ is shifted by 0.2-0.3 \kms\ from $\mu_{\rm n}$, regardless the method used. Similarly, $\sigma_{\rm b} > \sigma_{\mathrm n}$. 
The simplest explanation is that there is some warmer CNM along the line of sight that is less significant for absorption. If there is gas with somewhat different temperatures, the central velocity might be different too, also increasing $\sigma_{\rm b}$.

\subsubsection{Comparison with GHIGLS}

\begin{figure}[!t]
    \centering
    \includegraphics[width=\linewidth]{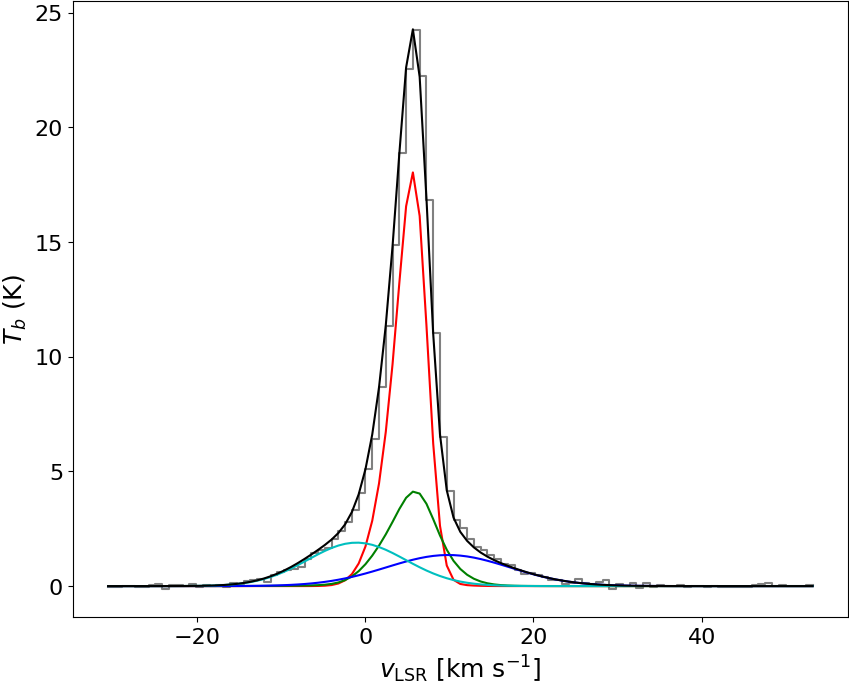}
    \caption{Gaussian decomposition by \ROHSA\ at the absorption position of DF from our final LVC model cube. The original signal is shown in gray and the total brightness temperature encoded by \ROHSA\ is shown in black. Colored lines represent each of the four components, as seen in Figure~\ref{fig:nh-std-maps}. Red represents CNM, Green: LNM, Blue: WNM$_{\rm A}$, Cyan: WNM$_{\rm D}$.
    }
    \label{fig:ROHSA_GHIGLS_DF_source_tmp}
\end{figure}

Figure~\ref{fig:ROHSA_GHIGLS_DF_source_tmp} shows, at the position of the absorption source, the 3-phase (four components) best model obtained with {\tt ROHSA} from averaging the three series presented in Section~\ref{subsec:uncertainties}. This spectrum, whose corresponding beam is that of GHIGLS (i.e., 9\farcm4), also shows the presence of unstable gas. 
Note that in this case, the model is constrained by the totality of the PPV cube and not only the emission at the position of the source (or from a small grid around). The resulting CNM component is still well described by a single Gaussian (although it shows a slight asymmetry toward negative velocities). This is compatible with what is seen in absorption. The corresponding Gaussian parameters are tabulated in Table~\ref{table:abs} (third row).

As for DHIGLS emission data, the presence of LNM lowers the brightness temperature of the cold gas at $\Trb=17.5$\,K, yield a lower spin temperature $\Trs=72$\,K, and a higher non-thermal broadening of the absorption line. From our best {\tt ROHSA} model, we find $f_{\rm CNM}=0.54\pm0.1$.

The variations in the physical quantities inferred from the three methods applied to GHIGLS and DHIGLS data reflect the difficulty of modeling the contribution of the CNM from the interpolated spectrum at the position of the absorption source. This translates to an uncertainty of the spin temperature of the gas, the non-thermal contribution to its velocity dispersion, and the CNM mass fraction along the line of sight (of similar amplitude to the uncertainty inferred from the stability analysis shown in Figure~\ref{fig:nh-std-maps} (top right panel)).

\subsection{Absorption against 4C +66.09}
\label{subsec:abs_UM}
Although not included in the field analyzed with {\tt ROHSA}, the DHIGLS survey contains another absorption line located in the NCPL (against Ursa Major), toward the background source 4C +66.09 (NVSS J094912 +661459), also shown by a white point in Figure~\ref{fig:T_b_full_cube_map}. The data from this emission-absorption pair are taken from the UM field, located east of DF.
Following the same methodology (i.e., interpolated spectrum and interpolated {\tt ROHSA} parameters maps), we find that the DHIGLS data are well fitted with two Gaussians (one broad and one narrow). The physical properties of the cold gas are similar to what was found in DF. For completeness, values are tabulated in the last two rows in Table~\ref{table:abs}.

\section{Discussion}
\label{sec:discussion}

\subsection{Mass fraction of CNM}
\label{sec:massfraccnm}

Based on coherent modeling of emission and absorption spectra distributed randomly over the entire intermediate to high latitude sky \citep{heiles_millennium_2003-1,murray2015,murray_2018,roy_temperature_2013,roy_temperature_2013-1}, it has been estimated that about 30\% of the \HI\ is in the cold CNM phase, and 20\% in the thermally unstable regime. \citet{marchal_2021a} found a similar fraction for the unstable gas in the NEP field of the GHIGLS survey but a much lower fraction of cold gas, about 8\%. 

By contrast, in the NCPL the mass fraction of CNM is significantly higher, on average 52\% in the loop.
We take this as an indicator that the dynamical process at the origin of the NCPL has been an effective trigger for the thermal phase transition and formation of the condensed CNM phase.

Similar mass fractions along the part of the loop  decomposed with {\tt ROHSA} (i.e., in Polaris and the Spider) suggest that the gas has been impacted in the much the same way.
However, there are significant column density differences along the loop. The Spider is a significant condensation that might reflect a pre-existing higher density region in the ISM being overrun by the dynamical process. In that context, the legs emanating from the Spider body might also relate to the response to the flow wrapping round this obstacle. 

Our best model indicates the presence of unstable gas in the loop, suggesting that the underlying dynamical process giving rise to the LNM phase is still in action. 
We note that the legs are traced in both high CNM mass fraction and the presence of LNM and there we do not immediately know in which direction the phase transition is occurring.

In the single UM line of sight in Ursa Major, a similarly high CNM mass fraction is observed, about 50\%, and the presence of LNM gas remains unclear. The comprehensive mapping of the thermal phases in Ursa Major needed for a comparative study of this part of the loop could be achieved using high resolution observations from the DHIGLS survey at 1'.



%




\subsection{Future work on the relationship to other tracers}
\label{sec:othertracers}


%
On large areas of the high-latitude sky, \citet{clark_2019} and \citet{murray_2020} have shown that the FIR emission from dust grains per H atom probed by \Planck\ increases monotonically with the fraction of cold gas inferred from the GALFA-HI survey \citep{peek_galfa-hi_2011,peek_galfa-h_2018}.
However, they could not discriminate between different scenarios that could underlie this effect:
the presence of molecular hydrogen yet no CO (CO dark gas), changes in dust emissivity, and/or changes in the dust-to-gas ratio associated with the thermal phase transition.  

In principle, our multiphase model offers such an opportunity to re-examine the FIR-\nh\ correlation in the NCPL.
For example, we find high CNM mass fractions in the body of the Spider (about 70\%) where the formation of CO was mapped and studied by \citet{barr2010}. There is excess emission relative to \nh\ and the dust is cooler \citep{planck2011-7.12,planck2013-p06b}. The interrelationships remain to be explored in more detail to understand what conditions can initiate a second phase transition to the molecular phase. In the Spider this would best be done by Gaussian decomposition of the DHIGLS data and Herschel imaging, both at higher angular resolution than GHIGLS and \Planck, respectively.


\begin{figure}[!t]
    \centering
    \includegraphics[width=\linewidth]{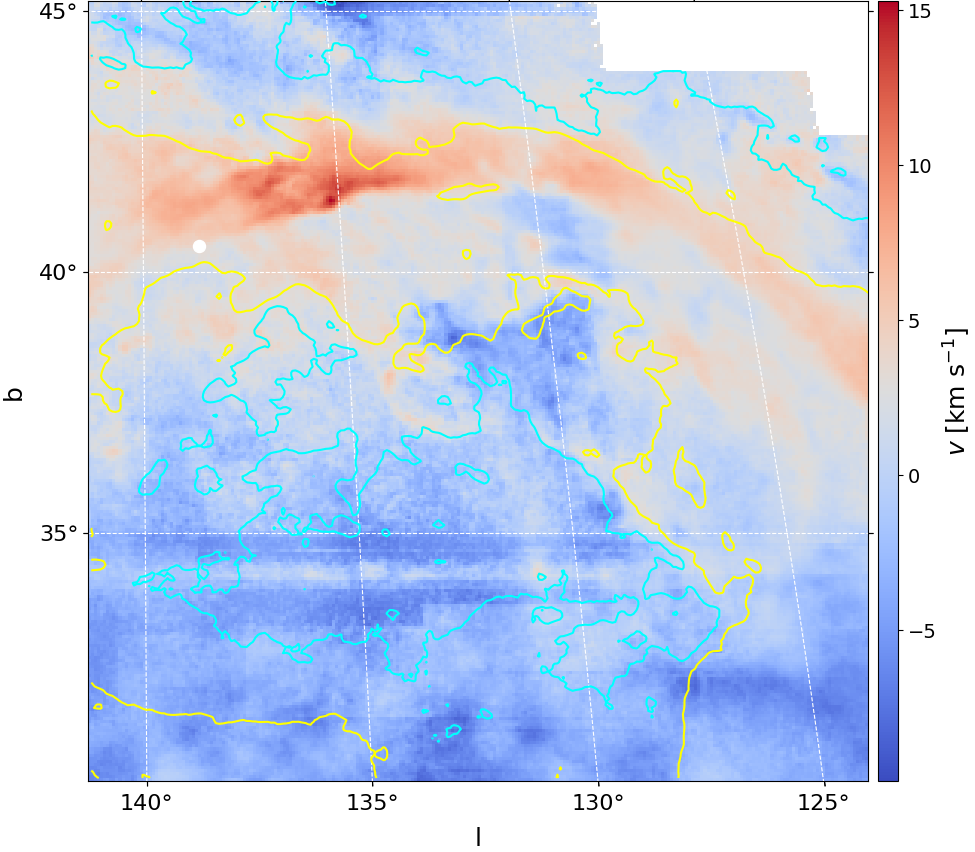}
    \caption{Centroid velocity map of WNM$_{\rm A}$+WNM$_{\rm D}$ from our best model inferred with {\tt ROHSA}. (The geometric/non-physical structure in regions of low intensity arises from imperfect stray radiation correction in blocks of the survey observed at different times.)
    Contours and annotation are the same as in Figure~\ref{fig:nh-std-maps}.
    }
    \label{fig:CV-WNMs}
\end{figure}

\subsection{Insight into the origin of the loop}
\label{sec:insightorigin}

From our Gaussian decomposition of the NCPL, we have concentrated on the cold/lukewarm gas following the loop. However, the warm gas, combining WNM$_{\rm D}$ and WNM$_{\rm A}$ components in a PPV cube, is also of interest.
Figure~\ref{fig:CV-WNMs} shows the velocity centroid of this gas. There are two features of note: the warm arch tracing the cooler loop as already seen in Figure~\ref{fig:nh-std-maps}, and the relative motion of the arch, which is related to the velocity gradient noted by  \citet{meyer91} and motivated their model of an expanding cylinder.
In a comprehensive study of a target cloud to the east of the region that we have analysed,  \citet{skalidis2021hi} conclude that overall the magnetic field
morphology is consistent with an expanding bubble or cylinder.

The remarkable spatial correspondence between this warm moving arch and the cold/lukewarm gas suggests that WNM$_{\rm A}$ is a relic of the large scale organized dynamical process that has triggered the phase transition in the loop.
The relationship to 3D dust extinction and the topology of extensions of the Local Bubble is discussed in more detail by \citet{mandm22}.

\section{Summary} 
\label{sec:summary}

Our novel study of the multi-phase properties of the North Celestial Pole Loop is based on \HI spectra from GHIGLS. 
We used \ROHSA\ to decompose the spectral data in emission to model their multiphase structure and corroborated this with limited absorption spectra that were available.
The main conclusions are as follows. 

\begin{itemize}
    \itemsep-0.2em
    \item A phase transition took place in the NCPL that led to the formation of a significant fraction of cold gas (about 50\% along the loop on average). Similar CNM mass fractions are observed in Polaris, the Spider and Ursa Major, suggesting a common origin of the dynamical process that triggered the thermal condensation.
    Our best model contains the presence of unstable gas which suggests that the dynamical process at the origin of this condensation is still in action.

    \item From absorption line measurement, we find that turbulence in the cold phase is similar in the Spider and Ursa Major and is supersonic with a turbulent sonic Mach number of about 3. This value is close to the peak of the histogram ($M_t\sim3$) of all 21-Sponge survey sources analyzed by \cite{murray2015}.

    \item We observe the presence of a well-defined warm arch associated with the loop, whose projected component of the velocity field is about 14\,\kms. The spatial correlation between this arch and the cold/lukewarm part of the loop suggests that it could be a relic of the large scale organized dynamical process that triggered the phase transition.
\end{itemize}

Extending the scope of this study are many possible future investigations.  For example, Zeeman measurements of the magnetic field strength by \citet{heiles_1989} could be correlated with the CNM mass fraction to investigate the influence of the magnetic field on dynamics and phase transitions, as can also be explored in numerical simulations such as by \citet{valdivia2016}.
Another possibility would be to 
study the physical properties of the \HI\ structures found by segementation of CNM, LNM, and WNM maps \citep{marchal_2021b}. This might best be done starting with a \ROHSA\ decomposition of the higher resolution DF data.

\acknowledgments
We acknowledge support from the Natural Sciences and Engineering Research Council (NSERC) of Canada. 
This work took place under the Summer Undergraduate Research Program (SURP), hosted by the University of Toronto. This research has also made use of the NASA Astrophysics Data System. We appreciate enlightening conversations with M.-A. Miville-Desch\^enes and F. Buckland-Willis. 
We thank the anonymous referee whose comments and suggestions have improved this manuscript.

\software{AstroPy, a community-developed core Python package for Astronomy \citep{astropy_2013, astropy_2018}, CMasher \cite{velden_2020}, NumPy \citep{van_der_walt_2011}, Matplotlib \citep{hunter_2007}, and, scikit-image \citep{scikit-image}, and SciPy \citep{SciPy-NMeth}.}

\clearpage
\appendix

\restartappendixnumbering

\section{In-painting}
\label{app:inpainting}

In directions where there is identifiable extra-galactic emission within the spectral range being analysed (including emission-free end channels), the \ROHSA\ decomposition modelling the Galactic emission can be affected, either ``wasting" Gaussians to fit this contaminating emission, or leaving large residuals. 

\begin{figure}[!ht]
    \centering
    \includegraphics[width=\linewidth]{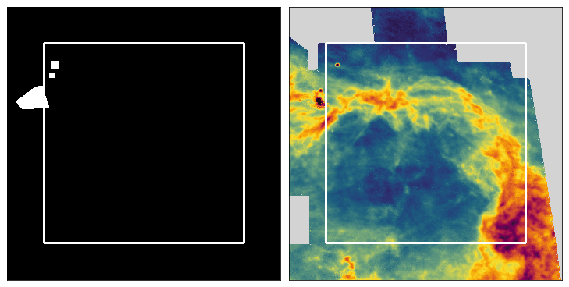}
    \caption{Polygons (white regions in left panel) were developed to isolate contaminating extra-galactic emission. This contamination can be seen in the map of the total column density (right) and is removed channel by channel by in-painting, with the final result seen in Figure~\ref{fig:T_b_full_cube_map}.
    Large white box outlines the sub-region of the data that we chose to focus on for the analysis.
    }
    \label{fig:inpaint}
\end{figure}

A main area of contamination is the large polygon seen in the left panel of Figure~\ref{fig:inpaint}.
For the modelling here of LVC or IVC emission, this polygon was defined conservatively to contain the emission of M81 \citep{adler1996} and other members of the group including M82, NGC 3077, and NGC 2976 near the west, north, and east apices, respectively, and the widespread tidally-disturbed gas \citep{yun1994,yun2000,chyn08,deblok2018,sorgho2019}. 

(This polygon also includes the GHIGLS-supplied mask of positions at which it was not possible to remove spectral baselines.) In cropping the region to be analysed here, we deliberately avoided most of this polygon, but it is unavoidable in the field studied by \cite{vujeva2022} toward Ursa Major to the east.

The two smaller polygons to the north contain emission near HIJASS~J1029+68 and IC~2574, respectively \citep{chyn09,sorgho2019}. 

These polygons were then in-painted channel by channel using built-in tool from SciPy: $\tt scipy.interpolate.griddata()$, adopting simple linear interpolation in spanning these sizeable polygons.
After the \ROHSA\ decomposition, these polygons are masked in further quantitative analysis.
\section{Maps of individual Gaussian components}
\label{app:maps}
\setcounter{figure}{0}

\begin{figure*}[!t]
    \centering
    \includegraphics[width=\linewidth]{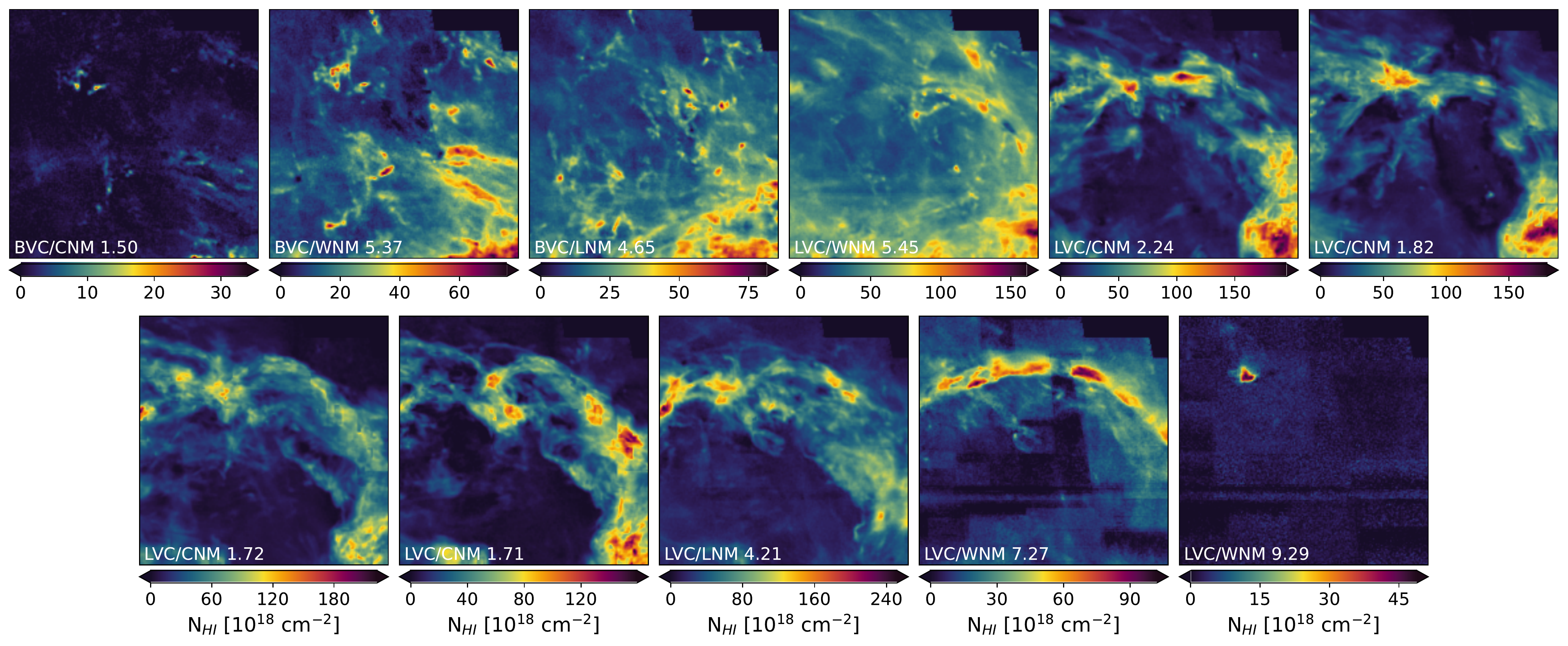}
    \caption{Mosaic of \nh\ maps of the individual Gaussian components. The maps are sorted by mean velocity. The labels in the bottom left corner represent the sorting of that particular Gaussian component. The number beside this label is the mean velocity dispersion ($\sigma$) in units of \kms. Note the different ranges of the color bars.}
    \label{fig:col-dens-mosaic}
\end{figure*}
\begin{figure*}[!t]
    \centering
    \includegraphics[width=\linewidth]{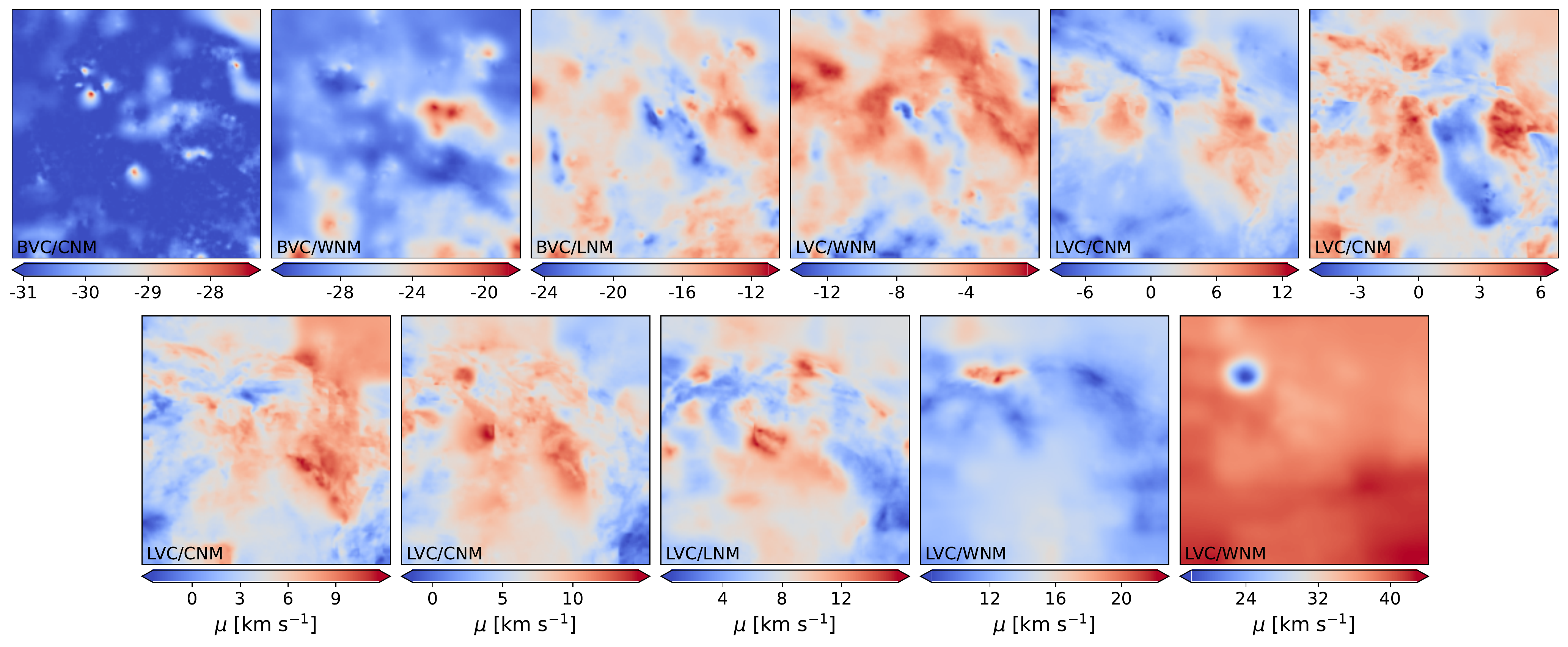}
    \caption{Mosaic like Figure~\ref{fig:col-dens-mosaic}, but of maps of the mean velocity, $\mu$, of the individual Gaussian components.}
    \label{fig:vel-mosaic}
\end{figure*}
\begin{figure*}[!t]
    \centering
    \includegraphics[width=\linewidth]{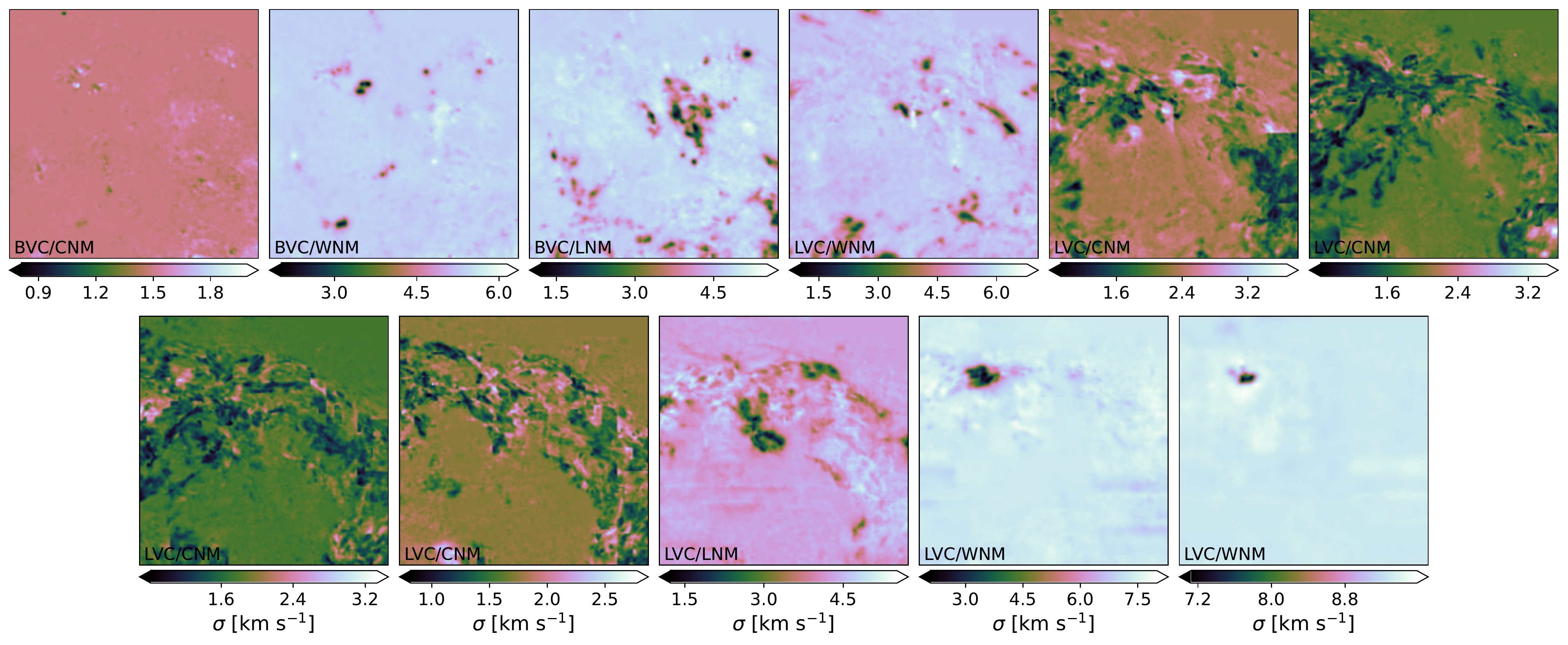}
    \caption{Mosaic like Figure~\ref{fig:col-dens-mosaic}, but of maps of the velocity dispersion, $\sigma$, of the individual Gaussian components.}
    \label{fig:sigma-mosaic}
\end{figure*}

Figure~\ref{fig:col-dens-mosaic} shows the column density maps of each Gaussian sorted by increasing column density-weighted mean velocities (see Table~\ref{table::mean_prop}). Figures~\ref{fig:vel-mosaic} and \ref{fig:sigma-mosaic} shows the corresponding velocity fields and dispersion velocity fields.
Note, the labels overlaid on the panels in Figure~\ref{fig:col-dens-mosaic} help to associate them with the clusters in Figure~\ref{fig:2d_hist}: e.g., for $G_7$ (row 2, column 1), ``LVC/CNM 1.72'' indicates the velocity range, the thermal phase, and the column density-weighted mean velocity dispersion $\left<\sigmab_n\right>$ in\,\kms\ as tabulated in Table~\ref{table::mean_var_NCPL}.

\clearpage
\bibliography{main}
\end{document}